\documentclass[a4paper,12pt]{article}
\pdfoutput=1 

\usepackage{jcappub} 

\usepackage[T1]{fontenc} 

\usepackage{tikz} 
\usepackage{graphicx}
\usepackage{subcaption}
\usepackage{amsfonts}
\usepackage{amsmath,amssymb}
\usepackage{hyperref}
\usepackage{units}
\usepackage{color}
\usepackage{xcolor}
\usepackage{dcolumn}
\usepackage{bm}
\usepackage{mathrsfs}
\usepackage{float}
\usepackage{cleveref}
\usepackage[normalem]{ulem}
\usepackage{appendix}
\usepackage{mathtools}
\usepackage{esvect}
\usepackage{array}
\usepackage{multirow}
\usepackage{empheq}

\newcommand{\be}{\begin{equation}}
\newcommand{\ee}{\end{equation}}
\newcommand{\beal}{\begin{aligned}}
\newcommand{\eeal}{\end{aligned}}
\newcommand{\bea}{\begin{eqnarray}}
\newcommand{\eea}{\end{eqnarray}}

\title{Ultra slow-roll with a black hole}

\author[a,1]{Lewis Croney,\note{Corresponding author.}}
\author[a,b]{Ruth Gregory,}
\author[a]{Sam Patrick}

\affiliation[a]{Department of Physics, King’s College London, University of London, Strand, London, WC2R 2LS, UK}
\affiliation[b]{Perimeter Institute, 31 Caroline Street North, Waterloo, ON, N2L 2Y5, Canada}

\emailAdd{lewis.croney@kcl.ac.uk}
\emailAdd{ruth.gregory@kcl.ac.uk}
\emailAdd{samuel\_christian.patrick@kcl.ac.uk}

\abstract{
We investigate ultra slow-roll inflation with a seed black hole in a de Sitter background.
By numerically tracking transitions from slow-roll to ultra slow-roll inflation, we find that quasi-normal mode solutions of the scalar field are excited following the decay of the slow-roll attractor, depending on the mass of the black hole.
For small black holes, the picture is similar to standard inflation with the usual damping of the scalar field; with a large black hole, we find that the ringing modes dominate. It is believed that the transition to ultra slow-roll in the pure inflationary case enhances the peak of the primordial power spectrum, thereby increasing the likelihood of primordial black hole formation. We comment on how the novel ringing behaviour due to the seed black hole might impact on cosmological perturbations.}

\begin{document}
\maketitle
\flushbottom

\section{Introduction}
\label{sec:intro}

The inflationary paradigm is very well established as a mechanism by which
the early universe emerges in a highly homogeneous and isotropic state \cite{Liddle:2000cg}. While originally intended to dilute exotic early universe phenomena, such as magnetic monopoles, and solve the horizon and flatness problems, inflation is also predictive in providing the seeds for structure formation.
The universe undergoes a period of accelerated expansion, typically driven 
by the vacuum energy of a scalar field, the {\it inflaton}. In {\it slow-roll} (SR)
inflation, the slope of the scalar potential is sufficiently mild such that the motion of the scalar is friction dominated. This allows the scalar to evolve slowly enough that its kinetic energy remains subdominant to its potential energy, which results in an approximate effective cosmological constant contribution to the Einstein equations. 
The consistency of this approach is encoded in the slow-roll parameters \cite{Liddle:1992wi}, which express conditions on the potential to support this motion.
More recently, ultra slow-roll (USR) inflation has become increasingly 
popular as a mechanism for enhancing the peak in the primordial power spectrum, which can in turn facilitate the formation of primordial black holes (PBH)
that can then contribute to the dark matter budget \cite{Carr:2016drx}.
In USR, the potential driving term is assumed negligible and instead the scalar exponentially decelerates from Hubble friction. While the first SR parameter $\epsilon \propto a^{-6}$ is indeed small in USR, the second SR parameter $\eta \approx -6$ is not. The amplitude of the curvature perturbation has a similar form as in SR \cite{Namjoo:2012aa}, $\mathcal{P_R} \propto \epsilon^{-1} \propto a^6$, which instead grows rapidly during USR, until the breakdown of perturbation theory.

Although it is usually assumed that inflation begins in a geometrically smooth
universe, there is always the possibility of an innate black hole
induced from the initial condition of the universe. 
Equally, while we do not fully understand the physics of the pre-inflationary era, it is entirely possible that black holes were formed, say from a first order phase transition \cite{Gregory:2020hia}. 
How does the inflationary cosmology
react to such black holes? The case of SR inflation was studied in \cite{Chadburn:2013mta, Gregory:2017sor, Gregory:2018ghc}
where it was found that scalar energy slowly accreted onto the black hole,
and the black hole dragged the scalar field in its vicinity. Although the black
hole affects the scalar evolution locally, far from the black hole the standard
SR approximation is restored. 
Although cosmological perturbation theory has not been calculated
for this spacetime, it seems likely that the impact of the black hole 
will be limited. However, for USR inflation, where the flatness of the
potential allows for larger fluctuations, any inherent spatial inhomogeneity
could have a more significant impact on scalar perturbation theory.

In this paper we examine the impact of a black hole on scalar evolution
when the potential enters a period of USR. We find that
whereas SR with a black hole proceeds analogously to the pure cosmological case,
USR can be very different. For SR, the scalar evolution is friction
dominated, and the effect of the black hole is simply to imprint a
spatial profile on top of this, but the feature of friction domination persists.
In USR however, the cosmological scalar evolution is a `free fall', with
a steadily decreasing kinetic energy. With the addition of a black hole, 
once the potential gradient driving term is switched off, the scalar 
will now ``ring down'', accessing the quasi-normal modes of the
cosmological black hole spacetime.
This means that there is not only spatial variation, but also time
variation imprinted on the background evolution.
In pure de Sitter, radial dependence can only be accessed at the level of the perturbations; instead with a black hole, these modes occur at the level of the background. This offers the prospect of modifying the enhancement of the peak of primordial power.

We first review the scalar cosmological equations in the presence of the black hole, in which a time coordinate $T$ is identified that tracks the scalar evolution. For SR with a black hole, we show that the scalar equation of motion reverts to the usual form far from the black hole.
Following the same procedure for USR however, we end up with exponentially damped scalar time dependence, rather than linear as in SR. This then suggests that we instead look for a more general radial and time dependence in our scalar, which leads to quasi-normal mode solutions.
Returning to de Sitter, we elucidate the role of the de Sitter quasi-normal modes for the scalar evolution.
Next we consider a transition from SR to USR, first analytically in de Sitter, and then numerically for Schwarzschild-de Sitter. We find that the attractor solution for SR subsequently decays into a linear combination of quasi-normal modes following USR entry, analogous to the ringdown of black holes. We conclude by commenting on the oscillatory quasi-normal modes arising at the level of the background for cosmological perturbation theory, and the potential impact on the peak of the primordial power spectrum.

\section{Slow-roll with a black hole}
\label{SRwithaBHSec}

We start by reviewing the impact of a black hole on a slowly rolling scalar field. 
Recall that in slow-roll inflation, the scalar field is spatially homogeneous,
and its motion is friction dominated. The slowly rolling scalar energy-momentum is then dominated
by its potential, leading to a close-to-cosmological-constant equation of state and an
accelerating universe. Maintaining this motion gives constraints on the scalar potential, 
encoded in the \emph{slow-roll parameters} that we will define presently.
The main challenge to including a black hole in this picture is the differing perspective
of a cosmological FRW universe, where the geometry is primarily time-dependent,
and a black hole universe, where the geometry is primarily space-dependent.
The trick to bridging this disparity is to identify a time coordinate upon which
the scalar depends, that naturally accommodates the black hole advanced time near
the black hole horizon, and cosmological time far from the black hole (and beyond the 
cosmological horizon in static coordinates).

Writing the scalar field as $\varphi$, and assuming minimal coupling to
Einstein gravity, the action for our system is,
\begin{equation} \label{action1}
\mathcal{S} = \int d^4 x\sqrt{-g} \left[ \frac{M_p^2\mathcal{R}}{2} 
- \frac{1}{2}(\nabla\varphi)^2 - W(\varphi)\right],
\end{equation}
where $\mathcal{R}$ is the Ricci scalar, $\nabla$ is the covariant derivative, 
$W(\varphi)$ is the potential of the scalar field, and $M_p = \sqrt{1/8\pi G}$ 
is the reduced Planck mass. The 
equations of motion are then obtained as,
\be\label{EinsteinScalarEM}
\beal
G_{\mu \nu} &= \frac{1}{M_p^2} T_{\mu \nu} 
= \frac{1}{M_p^2} \left [ \nabla_{\mu} \varphi \nabla_{\nu} \varphi - g_{\mu \nu} \left(\frac{1}{2} (\nabla \varphi)^2 + W(\varphi)\right) \right] \; , \\
\Box \varphi &= \frac{\partial W}{\partial \varphi}\;.
\eeal
\end{equation}
In standard SR inflation, we take the usual FRW metric,
\begin{equation} \label{FRWMetric}
ds^2 = -dT_c^2 + a(T_c)^2 (dR_c^2 + R_c^2(d\theta^2 + \sin^2\theta \, d\phi^2)).
\end{equation}
and our equations of motion \eqref{EinsteinScalarEM} become,
\be
\label{FRWeqns}
\beal
H^2 = \left ( \frac{\dot a}{a} \right)^2 &= \frac{1}{3M_p^2} \left [ 
W(\varphi) + \frac12 {\dot\varphi}^2 \right ], \\
\ddot{\varphi}+ 3 H \dot{\varphi} &= - \frac{\partial W}{\partial \varphi}.
\eeal
\ee
The SR approximation then takes $\dot{\varphi}^2 \ll W(\varphi)$, so
that the Friedmann equation is dominated by the vacuum energy of the scalar,
and $\ddot{\varphi} \ll 3H\dot{\varphi}$ to maintain SR. These conditions
are encapsulated in the \emph{slow-roll parameters} \cite{Liddle:1992wi},
\be\label{SRBHParams}
\epsilon = \frac{M_p^2}{2} \left(\frac{W'}{W}\right)^2, 
\quad \eta = M_p^2 \frac{W''}{W},
\end{equation}
that ensure that the scalar field motion remains friction dominated, and
the expansion of the universe is dominated by vacuum energy.

To include a black hole in this picture, one has to accommodate the fact
that the natural coordinates for a black hole in a cosmological constant 
or de Sitter (dS) universe are `static',
\begin{equation} \label{BackgroundSdSMetric}
ds^2 = -f(r) dt^2 + \frac{1}{f(r)} dr^2 + r^2 (d\theta^2 + \sin^2\theta \, d\phi^2),
\end{equation}
with,
\begin{equation} \label{Backgroundfr}
f(r) = 1 - \frac{2GM}{r} - H^2 r^2,
\end{equation}
whereas the FRW metric is time dependent.
Here, $M$ is interpreted as the black hole mass \cite{Ghezelbash:2001vs}, 
and $H$ is the Hubble parameter, related to the cosmological constant 
via $\Lambda = 3H^2$.
For $GM \sqrt{\Lambda} < 1/3$, this spacetime has two physical horizons, and $r_b$ and $r_c$ denote the radii of the black hole and cosmological horizons
respectively\footnote{The case of $GM\sqrt{\Lambda} = 1/3$, the extremal limit where the horizons coincide, is called the Nariai limit \cite{1950SRToh..34..160N}.}.
In the absence of a black hole, it is straightforward to transform between the different
coordinate systems for dS, although adding
the black hole leads to a more complicated transformation.
Once one has an evolving scalar field sourcing the vacuum energy, the interplay
between the time dependence and the local black hole horizon is less well understood
(see \cite{Carrera:2008pi} for an overview of the issues). 

Using the intuition of the slow-roll scalar motion being dominated by friction, 
\cite{Chadburn:2013mta, Gregory:2017sor, Gregory:2018ghc} approached the
``slow-roll with a black hole'' set-up by noting that if the rate of change in vacuum 
energy is sufficiently slow, the spacetime can be approximated by 
a fixed Schwarzschild de Sitter (SdS) universe \eqref{BackgroundSdSMetric}, on which 
the scalar rolls. The problem is then dealt with by a back-reaction method, first 
finding the scalar solution on the background, then computing its gravitational effect.
Using the nature of friction dominated motion, the method is to find a timelike 
coordinate, $T_{\textrm{SR}}$, such that $\varphi = \varphi(T_{\textrm{SR}})$, 
and the equation of motion reduces to the SR friction dominated equation. 
Writing $T_{\textrm{SR}}=t + h(r)$, this is readily obtained as,
\be \label{SREoMBH}
\frac{\partial W}{\partial \varphi} = \Box \varphi =
\frac{1}{r^2} \left [ r^2 f(r) h'(r) \right] ' \dot{\varphi}(T_{\textrm{SR}})
- \frac{1-h'(r)^2}{f(r)} \; \ddot{\varphi}(T_{\textrm{SR}}),
\ee
where the last (acceleration) term is neglected for slow-roll. Since $\partial W/\partial\varphi$ is
purely a function of $T_{\textrm{SR}}$, we deduce that,
\be
\frac{1}{r^2} \left [ r^2 f(r) h'(r) \right] ' = -3\gamma_{\textrm{SR}},
\label{slowrollgamma}
\ee
so that the ``slow-roll'' scalar equation is,
\be
3 \gamma_{\textrm{SR}} \dot{\varphi} = - \frac{\partial W}{\partial\varphi},
\label{SReqn}
\ee
where $\gamma_{\textrm{SR}}$ is a constant determining the friction of the motion,
and the factor of 3 is placed so that in the absence of a black hole $\gamma_{\textrm{SR}}=H$,
and the scalar equation reduces to the standard slow-roll expression. 
$h(r)$ can then be found by integrating \eqref{slowrollgamma},
\be
h'(r) = \frac{(\beta_{\textrm{SR}} - \gamma_{\textrm{SR}} r^3)}{r^2f(r)},
\ee
then demanding regularity of the scalar on each of the black hole and 
cosmological horizons determines the constants,
\begin{equation} \label{betagammaSR}
\beta_{\textrm{SR}} = \frac{r_b^2 r_c^2 (r_b + r_c)}{r_c^3 - r_b^3}, 
\quad \gamma_{\textrm{SR}} = \frac{r_c^2 + r_b^2}{r_c^3 - r_b^3}.
\end{equation}
It is now easy to see that as the black hole radius $r_b\to0$, 
$\beta_{\textrm{SR}}\to0$ and $\gamma_{\textrm{SR}}\to H$ as claimed above.
In full, $T_{\textrm{SR}}$ has the rather lengthy form that nonetheless 
reduces to the local horizon Eddington-Finkelstein time at $r_b$ and $r_c$ (discussed in Appendix \ref{Penroseapp}), to ensure regularity,
\be
\beal
T_{\textrm{SR}} &= t - \frac{1}{2\kappa _c } \ln \left ( \frac{r_c -r}{r_c} \right ) 
+ \frac{1}{2\kappa _b } \ln \left ( \frac{r -r_b}{r_b } \right) \\
& \quad + \left ( \frac{r_c}{ 2\kappa_b r_b} -\frac{1}{2\kappa_n} \right ) 
\ln \left ( \frac{r -r_n}{ r_n } \right ) + \frac{r_b r_c }{ r_c -r_b }  \ln\frac{r}{r_0 } ,
\eeal
\label{SlowRollBHTime}
\ee
where $r_n =- (r_b +r_c )$ is the negative root of the SdS metric function $f(r)=0$, 
and $\kappa_i = f^\prime (r_i)/2$ are the surface gravities evaluated at the roots of $f(r)=0$, 
with $r_0$ being an arbitrary constant of integration. 

Having found this dependence for a scalar rolling on a SdS background, 
the energy-momentum of this source can be used to correct the gravitational
background \cite{Gregory:2018ghc} and find, to leading order, the 
geometry of the slowly-rolling scalar with a black hole (see also \cite{Beyen:2023rca}
for a more careful analysis).

This solution, as written above, is presented in terms of the static patch of SdS, 
and shows how the black hole retards the motion of the scalar in its vicinity.
From a cosmological perspective however we would not expect a black hole
to have a significant impact on the scalar rolling far from the black hole. To
check this, we have to look beyond the cosmological horizon. Since the slow-roll
time is a regular coordinate at $r_c$, we can extend beyond the cosmological 
horizon by, for example, defining an orthogonal 
basis via,
\be
dT_{\textrm{SR}} = dt + h' dr, \qquad
\frac{dR_{\textrm{SR}}}{R_{\textrm{SR}}} = dt + \frac{dr}{f^2 h'}.
\ee
Far from the black hole, the time component of this metric,
\be
g_{_{T_{_\textrm{SR}}T_{_\textrm{SR}}}} \sim \frac{H^2}{\gamma_{\textrm{SR}}^2} 
\left [1 + {\cal O} (r^{-2}) \right],
\ee
tends to a constant, and one can identify a coordinate analogous to the FRW time 
$T_c = H T_{\textrm{SR}}/\gamma_{\textrm{SR}}$ which, when substituted 
in \eqref{SReqn}, gives the canonical SR scalar equation,
\be
3 H \frac{d \varphi}{d T_c} = - \frac{\partial W}{\partial \varphi}.
\ee

\section{Ultra slow-roll inflation}

In standard slow-roll (SR) inflation, the dominant term in the scalar equation is the 
gradient of the scalar potential, thus the $\ddot{\varphi}$ term in the scalar
equation of motion is neglected leading to a friction dominated motion.
However, in ultra slow-roll (USR) inflation, the scalar enters an extremely flat range of the potential, so that $\partial W/\partial \varphi \approx 0$, and
the scalar acceleration becomes the relevant term in the equation of motion,
\begin{equation} \label{ultraslowrollpuredS}
\ddot{\varphi} + 3 H \dot{\varphi} \approx 0.
\end{equation}
The ``\textit{ultra}'' nomenclature is to signify the rapid deceleration of the scalar 
that will occur during USR.
Models of inflation with a period of USR have been of recent interest for giving 
rise to an enhanced peak in the primordial power spectrum, which could facilitate 
primordial black hole (PBH) production \cite{Germani:2017bcs, Jackson:2023obv}. 
This is exciting, since PBHs could constitute a large fraction of dark matter 
\cite{Carr:2016drx, Green:2020jor}.
The period of USR should not last too long, otherwise the scalar will 
decelerate to a halt, leading to eternal inflation \cite{Byrnes:2021jka}. 
Instead, the USR period should give way to another regime in the potential, 
such as another ordinary slow-roll period, as in some inflection point models \cite{Byrnes:2018txb}.
Furthermore, if the potential has \textit{any} non-zero $W'(\varphi)$, then a 
regime of USR is guaranteed to end once the scalar decelerates down to the 
point where the acceleration and potential derivative terms compete \cite{Dimopoulos:2017ged}.

Our aim is to now discover how a black hole modifies the pure cosmological 
USR scalar evolution. Note that the Hubble parameter $H$ remains roughly constant over the 
timescale of variation of the scalar since the Friedmann equations are dominated 
by an effective cosmological constant induced from the scalar vacuum energy,
$\Lambda = W(\varphi)/M_p^2$ (which remains constant due to the flat potential
in the USR regime).
We therefore take a ``test-field'' approximation, in which we analyse the scalar
motion on the SdS background. 

First, note that for a pure cosmological background \eqref{ultraslowrollpuredS}
implies that $\dot{\varphi} \propto e^{-3\sqrt{\Lambda}T}$, where we have
substituted for the scale factor from solving the Friedmann equation.
This means that the scalar slows down coherently and rapidly, with a direct
dependence (purely) on cosmological time. For a small black hole, we expect 
a very similar behaviour, but with a spatially dependent profile due to
the black hole `dragging' the scalar at its horizon. 

Just as \cite{Gregory:2018ghc}
took the SR approximation to mean neglecting the second derivatives of the
scalar field, here we take the defining feature of USR to mean neglecting the
gradient of the potential, thus the scalar equation of motion reduces to,
\be \label{ScalarUSR}
g^{\mu \nu} \nabla_{\mu} \nabla_{\nu} \varphi \approx 0.
\ee
For a spherically symmetric background, the minimal response of the 
scalar to the black hole will be to have a profile dependent on both time
and a radial coordinate. Additional angular dependence could arise, but 
this would be at the level of perturbation theory, as the background 
will not initiate this dependence. 

Note that we are looking for solutions to a wave equation that are regular 
at both the black hole and cosmological future event horizons.  Thinking about
the motion of the scalar field as it transitions from SR to USR, there will be a
spacelike hypersurface on which $\varphi = \varphi_0$, the value of the scalar
at which the transition occurs. The scalar will have some `momentum' that we
expect to decay as the Hubble friction starts to dominate over the negligible potential gradient. Where this hypersurface intersects
with each event horizon, the momentum of the scalar will be outgoing across
the horizon, thus we are looking for a solution to \eqref{ScalarUSR} that has no 
incoming component across either horizon.

This is a very familiar set-up in black hole physics, as clearly a solution with
these boundary conditions will decay in time, and corresponds to the ringdown 
of a black hole. The solutions corresponding to ringdown are known as
quasi-normal modes (QNMs) \cite{Berti:2009kk} and typically take the form,
\begin{equation} \label{QNMConvention}
\varphi = e^{-i \omega t} z(r),
\end{equation}
where the frequency, $\omega$ is complex $\omega = \omega_R + i \, \omega_I$
-- the (negative) imaginary part corresponding to the decay of the signal,
and the real part corresponding to the `pitch' of the ringing.
Quasi-normal modes play a significant role in black hole spacetimes, since they 
dominate the ringdown signal observed in the gravitational wave emission of binary 
black hole mergers, allowing identification of the parameters of the final black hole 
state \cite{LIGOScientific:2016aoc, Berti:2009kk}.

Thus, the individual solutions of the USR equation have a parallel with the QNM solutions
of an SdS, or indeed a dS background. This correspondence does not appear to have 
been noted in the existing ultra slow-roll literature, most likely because the cosmological
solution for the scalar depends solely on the FRW time, with spatial dependence only
introduced at the level of perturbation theory. However this correspondence could have 
been anticipated, since QNMs and USR are derived from the same equation, 
with equivalent boundary conditions. The key issue as to which QNMs are excited therefore 
comes down to the initial conditions of the scalar field as it exits slow-roll. 
For the cosmological background, the scalar is homogeneous, so only a QNM
with no spatial dependence is relevant. For the black hole background however, 
there is a radial dependence of the scalar, so we must be more careful about
the analysis, and anticipate that ``ringing'' QNMs can become relevant.

\subsection{The QNM analysis} \label{AnalyticdSQNMResultsSec}

In static coordinates, using \eqref{QNMConvention} for the form of the scalar field, from \eqref{ScalarUSR}
we obtain the eigenvalue equation,
\begin{equation} \label{SdSRadialProfileEq}
z'' + \left(\frac{f'}{f} + \frac{2}{r} \right) z' + \frac{\omega^2}{f^2}z = 0,
\end{equation}
which must be solved for the radial profile $z(r)$ with the appropriate
boundary conditions, namely that $\varphi$ is a function of the relevant 
Eddington-Finkelstein coordinate  at each horizon (discussed, along with the tortoise coordinate $r^*$, in Appendix \ref{Penroseapp}),
\begin{equation} \label{SdSQNMBCs}
\begin{aligned}
\varphi & \sim A_1 e^{-i\omega (t+r^*)} \;\;
&\Rightarrow \quad z(r) \;\; 
& \sim A_1 e^{-i\omega r^*} \quad &\textrm{ as } r \rightarrow r_b, \\ 
\quad \varphi & \sim e^{-i\omega(t-r^*)}  
& \Rightarrow \quad z(r) \;\; 
& \sim e^{i\omega r^*} \quad &\textrm{ as } r \rightarrow r_c, 
\end{aligned}
\end{equation}
for some constant $A_1$. Since quasi-normal modes are defined up to an overall 
multiplicative constant, we have used this freedom to fix the arbitrary constant 
on the outgoing boundary condition to unity.  

We can build intuition by considering the problem for pure de Sitter spacetime. 
The quasi-normal modes of de Sitter are known \cite{Lopez-Ortega:2006aal}, 
however we will be able to find more explicit radial profiles (working in the $l = 0$ case), 
which to our knowledge have not appeared in the existing literature. 
For pure de Sitter, the absence of the black hole horizon alters the boundary conditions.
Instead, in order to ensure the requirement of regularity at the boundaries 
(which are now the origin $r = 0$ and the horizon $r = 1/H$), we need,
\begin{equation} \label{zdSBC}
z(r) \textrm{ regular as } r \rightarrow 0, \quad z(r) \sim \exp{(i \omega \, r^{*})}
\textrm{ as } r \rightarrow 1/H.
\end{equation}
Since $f(r) = 1 - H^2 r^2$, the radial profile equation \eqref{SdSRadialProfileEq} explicitly becomes,
\begin{equation} \label{dSRadialProfileEq}
z'' + \left(\frac{2}{r} - \frac{2H^2r}{1-H^2r^2} \right) z' + \frac{\omega^2}{(1-H^2r^2)^2}z = 0.
\end{equation}

The time that the usual cosmologically slow (or ultra-slow) rolling scalar follows can be found 
by transforming the cosmological time from FRW to static coordinates (given in \eqref{dScoordstrafo}),
\begin{equation}
\varphi \propto e^{-3HT} = e^{-3Ht} (1-H^2 r^2)^{-3/2},
\end{equation}
which can be verified as a solution to \eqref{dSRadialProfileEq}.
For this solution we see that,
\begin{equation} \label{dSFRWexpsol}
\omega = - 3 H i, \quad z(r) = (1-H^2 r^2)^{-3/2},
\end{equation}
i.e.\ there is no oscillation, just a purely damped mode, with a radial dependence
on the static areal coordinate. Note that for pure de Sitter, this cosmological time coordinate $T$ that the scalar follows in USR, is identical to
the time coordinate that the scalar follows in SR, obtained from \eqref{SlowRollBHTime}
by setting $r_b\to0$.

However, this is not the only possible solution for ultra slow-roll, writing $z = \xi/r$, $x = Hr$, and $s = \omega/H$,
the equation of motion \eqref{dSRadialProfileEq} becomes of Legendre form,
\begin{equation} \label{LegendredS}
(1 - x^2) \frac{d^2\xi}{dx^2} - 2 x \frac{d\xi}{dx} + \left(2 + \frac{s^2}{1-x^2} \right) \xi = 0,
\end{equation}
with general solution $P^{i s}_{1}(x)$ and $Q^{i s}_{1}(x)$.

For $s \in \{-2i, -3i, -4i, ...\}$, it turns out that $P^{i s}_{1}(r)\equiv 0$,
and the usual techniques for finding degenerate linearly independent solutions
to the eigenvalue equation apply. Correcting appropriately for this leads to the 
solutions for the scalar field,
\begin{equation}
\begin{aligned}
\varphi_n & = e^{- n H t} \frac{1}{(1 - n )Hr} \left(e^{n H r^*} (H r - n) + e^{- n H r^*} (H r + n)\right) \\
& = e^{-n H u} \frac{1}{(1 -n)Hr} \left(Hr - n 
+ \left(\frac{1-Hr}{1+Hr}\right)^{n} (Hr +n)\right),
\end{aligned}
\label{varphi-n}
\end{equation}
where $n \in \{2, 3, 4, ...\}$, and the second line explicitly extracts the desired $e^{-i \omega u}$ 
outgoing boundary condition \eqref{SdSQNMBCs} at the cosmological horizon, with QNM frequencies $\omega \in \{-2Hi, -3Hi, -4Hi, ...\}$.
As noted previously, these frequencies were known \cite{Lopez-Ortega:2006aal}, 
however the corresponding radial profiles were only given in terms of 
hypergeometric functions, perhaps because authors are usually more interested 
in general results, such as for general multipole $l$ and general dimensions $D$ 
\cite{Lopez-Ortega:2006aal, Du:2004jt}. Instead here we have obtained explicit 
expressions for the radial profile. 
In pure de Sitter, since the frequencies have vanishing real 
part, these modes are purely damped in time, with no oscillatory behaviour.

In FRW coordinates, these solutions can be written as,
\be
\varphi_n = \frac{(H R_c \, e^{H T_c} - n)(1+ H R_c \, e^{H T_c})^n
+ (H R_c \, e^{H T_c} + n)(1- H R_c \, e^{H T_c})^n}{(1-n) H R_c \, e^{(n+1) H T_c}} ,
\label{phisolcos}
\ee
and indeed $n=3$ is the only solution that is spatially homogeneous in these coordinates.

\subsection{Numerical results in SdS} 
\label{QNMSpectralNumResults}

The quasi-normal modes of the Schwarzschild metric are not known analytically, 
but instead require numerical calculation \cite{Cardoso:2003vt}. Therefore, we 
expect that in the more general Schwarzschild-de Sitter case, a numerical approach 
will also be required\footnote{However, analytic results can be obtained in the special 
case of the Nariai limit \cite{Cardoso:2003sw}.}. Several numerical methods for 
calculating QNMs exist, including methods based on WKB approximations 
\cite{Schutz:1985km, Konoplya:2019hlu}. A sixth order WKB method was 
applied to find QNM frequencies for SdS in \cite{Zhidenko:2003wq}. 
However, this approach missed a set of purely imaginary modes (which 
effectively extend from the pure de Sitter modes). These additional modes 
(and the original set) were identified via pseudospectral methods in 
\cite{Jansen:2017oag}, where the author developed the Mathematica 
package ~QNMSpectral \cite{Jansen:2017oag} to compute quasi-normal 
mode frequencies and radial profiles. This package has a wider range of 
applicability than merely Schwarzschild-de Sitter, being additionally able 
to handle flat/AdS asymptotics, and coupled equations. The main difficulty 
in its application comes from writing the equations of motion (for the 
perturbation, in our case a scalar $\varphi$) in a form that the code is 
prepared to accept. For our purposes in SdS, the equation to input into 
QNMSpectral is given in an appendix of \cite{Jansen:2017oag}, however 
there are coefficient typos, which are corrected in an appendix of \cite{Jansen:2018fjd}. 

Many of the numerical results and observations we discuss here are also 
provided in the original QNMSpectral reference \cite{Jansen:2017oag}, 
but they are of sufficient importance that we include them here for completeness. 
We also mostly follow their conventions for how such results are presented. 
Additional results on the radial profiles and the real parts of QNM frequencies 
are provided there, but omitted here.

In Schwarzschild-de Sitter systems, there are two independent length scales at play. 
Comparing the black hole mass $M$ against $L \equiv \sqrt{3/\Lambda}$ (the de Sitter curvature scale in the absence of the black hole)
provides one way of identifying the relative size of the black hole horizon against that 
of the cosmological horizon, although the cosmological horizon will decrease from $L^{-1}$
as we increase $M$. We choose to fix $L=1$ and vary the black hole mass $M$, all the 
way from no black hole ($GM/L = 0$) to the Nariai limit ($GM/L = \frac{1}{3\sqrt{3}}$) where 
the horizons coincide\footnote{We will only go \textit{near} to the Nariai limit, since 
this special case requires additional care \cite{Cardoso:2003sw}.}.
Any other setup of $M$ and $L$ with the same $GM/L$ will be physically equivalent. 

In the case of pure de Sitter, we calculated the quasi-normal mode frequencies 
analytically, and found that there are only purely imaginary modes, with a constant 
spacing between successive modes, \eqref{varphi-n}. 
As soon as one switches on the black hole 
mass $M$, there are additional modes that appear with a real part (which gives 
rise to oscillatory ringing). The purely imaginary modes from de Sitter however also
remain, and stay purely imaginary as we increase $M$, throughout the entire range of $GM/L$.
If $GM/L$ is sufficiently small (compared to the Nariai limit $\frac{1}{3\sqrt{3}}$), 
then there is effectively a tower of modes that are very close to that of pure de Sitter 
(with de Sitter horizon radius $L$), and a tower of modes that are very close to that of pure 
Schwarzschild (with Schwarzschild mass $M$). This is expected, since in this case 
the horizons are well-separated.
Henceforth, we will refer to these two towers as the \textit{de Sitter branch} and 
\textit{Schwarzschild branch} of quasi-normal modes.

\begin{figure}[ht]
\centering
\includegraphics[width=0.66\textwidth]{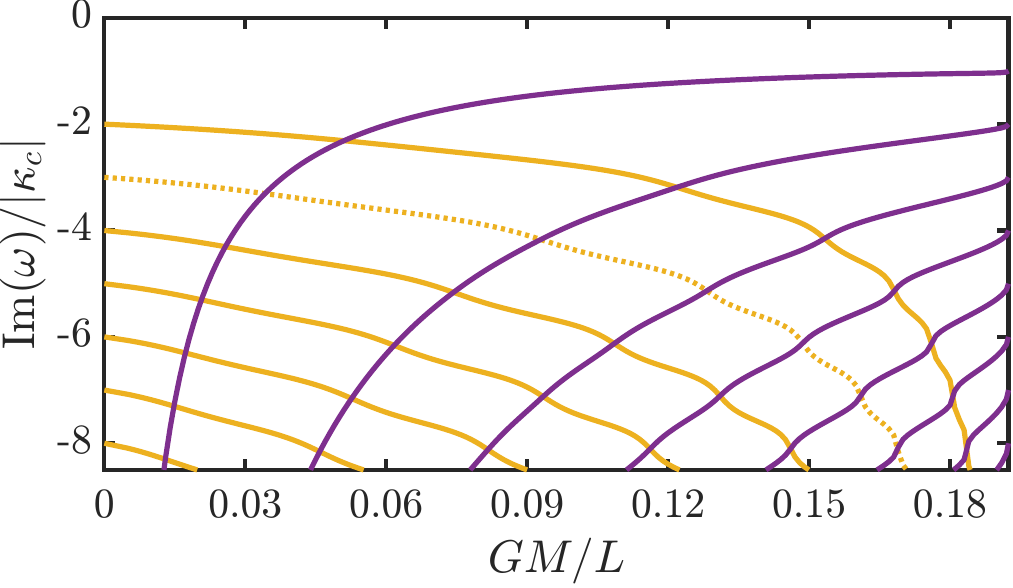}
\caption{Imaginary parts of the longest lived quasi-normal mode frequencies in Schwarzschild-de Sitter. The orange curves are modes from the de Sitter branch, and the purple curves are modes from the Schwarzschild branch. We highlight that 2nd de Sitter mode (which is homogeneous in FRW coordinates for pure de Sitter) as a dotted curve. For lower $GM/L$, the de Sitter modes are longer lived, until a crossover around $GM/L = 0.0515$, when the 1st Schwarzschild mode takes over. Produced using output from QNMSpectral, in the spirit of an equivalent figure originally from \cite{Jansen:2017oag}.}
\label{ImagFreqPlot}
\end{figure}

In Figure \ref{ImagFreqPlot}, we show the imaginary parts of the frequencies for the 
two branches of QNMs, normalised by the surface gravity of the cosmological 
horizon $|\kappa_c|$. The imaginary part of a QNM frequency gives a notion of the 
lifetime of that mode; a smaller imaginary part implies that the mode decays away 
more slowly in time. In orange we show the de Sitter branch modes, and in purple we show the 
Schwarzschild branch modes. This is in the spirit of a similar figure 
presented in \cite{Jansen:2017oag}.
Note that we only show a finite number of modes, since the QNMSpectral 
implementation can only resolve so many within a finite pseudospectral calculation, 
but there are in fact an infinite tower of modes within each branch for every 
$0 < GM/L < \frac{1}{3\sqrt{3}}$.
Looking at Figure \ref{ImagFreqPlot}, for $GM/L = 0$, we see the QNM frequencies 
that we identified for de Sitter. As we increase $GM/L$, we see that the constant 
spacing between de Sitter branch modes fades away, and we also see the 
emergence of the Schwarzschild modes. Broadly speaking, the de Sitter modes 
become more short-lived, and the Schwarzschild modes become more long-lived 
(except for oscillations closer to the Nariai limit). We see that for sufficiently small 
$GM/L$, the longest lived mode belongs to the de Sitter branch, however around 
$GM/L = 0.0515$, this switches to become one from the Schwarzschild branch, 
which persists up to the Nariai limit. 

Since the imaginary part of QNM frequencies for each branch do not cross each 
other as $GM/L$ varies, we can refer to modes within each branch as the \textit{1st mode}, 
\textit{2nd mode}, and so on. We highlight that the mode that originates from homogeneity 
in FRW coordinates \eqref{FRWMetric} belongs to the \textit{2nd de Sitter} mode, 
having $\omega/|\kappa_c| = -3$ in the pure de Sitter case.

Having outlined how the QNMs for SdS are computed, we now turn to the application
of these functions in determining transitions of the scalar from a SR to USR regime in 
the presence of a black hole.

\section{SR to USR transitions} \label{TransitionSec}

Now that we understand scalar solutions around black holes in both the 
SR and USR regimes, we now turn to the transition between them. 
This transition is key for giving rise to primordial power spectrum peak 
enhancement, and hence production of PBHs. Any potential enhancement
of such perturbations is therefore highly relevant.

\begin{figure}[ht]
\centering
\includegraphics[width=0.66\textwidth]{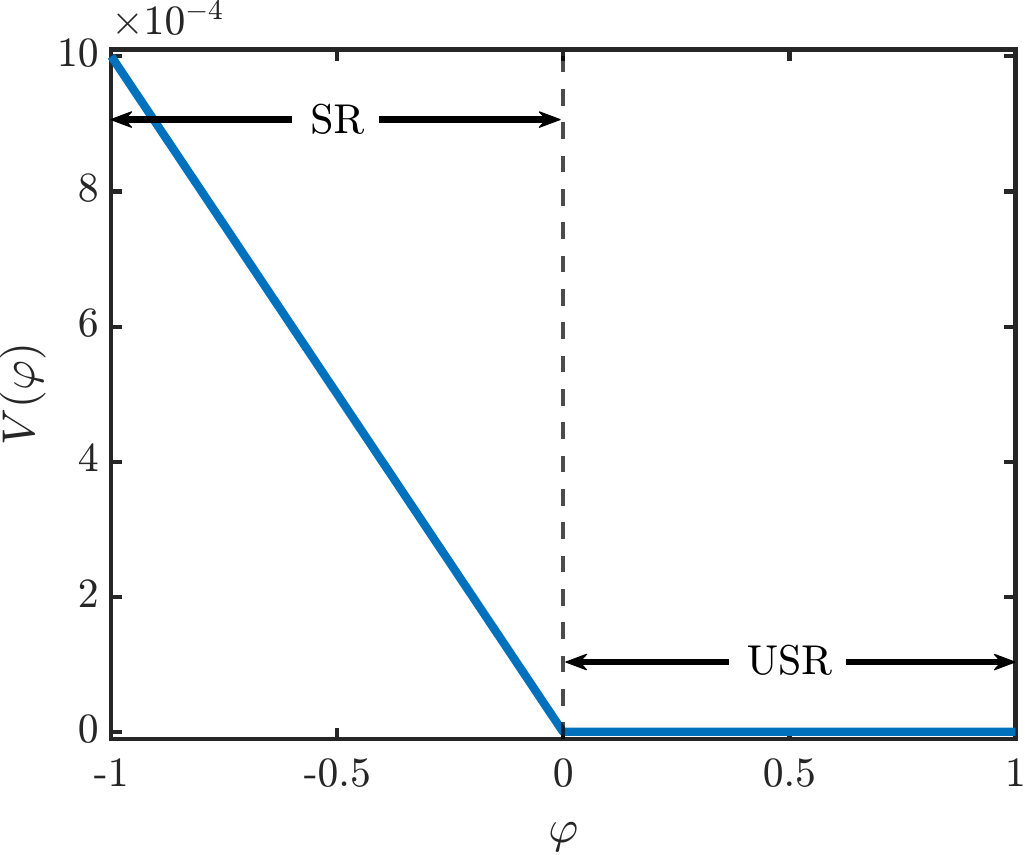}
\caption{A prototypical potential for $V(\varphi) \equiv W(\varphi) - M_p^2\Lambda$ for a slow-roll to ultra slow-roll transition. We work in units where $M_p = 1$ for brevity. The SR and USR ranges of the potential are highlighted, and delineated by a dotted vertical line. Since we set $L = 1$, we have $\Lambda = 3$. Note that the potential ends with $V(\varphi) = 0$; this means that we are using the background value of $\Lambda$ as the USR value of the overall potential $W(\varphi)$. In the slow-roll regime, this potential satisfies the SR conditions outlined in \eqref{SRBHParams}, since the maximum of $V(\varphi)$ is $10^{-3}$ here (in units with $M_p = 1)$. A caveat about this relating to initial conditions is discussed in Appendix \ref{NumFitApp}.}
\label{PotentialExamplePlot}
\end{figure}

To extract the phenomenology of USR with an existing black hole, we 
consider the simple scalar potential shown in Figure \ref{PotentialExamplePlot},
$V(\varphi) \equiv W(\varphi) - M_p^2\Lambda$, where $\Lambda$ is 
fixed by the vacuum energy that the scalar experiences while rolling in the USR (flat) range of the potential. We also restrict $V(\varphi) \ll M_p^2\Lambda$, 
so that the background metric can be treated as fixed throughout the transition. We discuss a subtlety regarding initial conditions for this setup in Appendix \ref{NumFitApp}.
For convenience we have set the transition between SR and USR at $\varphi = 0$, 
so we can write,
\begin{equation}
V(\varphi) =
\begin{cases}
W'_0 \; \varphi & \text{if } \varphi < 0 \\
0 & \text{if } \varphi \geq 0
\end{cases},
\label{PotentialFunction}
\end{equation}
where $W'_0$ represents an approximately negative potential gradient towards 
the end of the SR phase. The particular case of $W'_0$ we use for our numerical 
results here is $- 10^{-3}$ (in units where $M_p = 1$), but provided $W'_0$ is sufficiently small to allow the 
background metric to be fixed, the results will scale appropriately for different $W'_0$.

This model of the potential is simplistic, but should suffice for modelling the 
SR-USR transition. 
Although there is an instant transition between the SR (sloped) and USR (flat) range of the potential, 
a smooth transition between these two regimes would introduce unnecessary additional
complexity into the numerics. 
Again for simplicity, since we want to focus on the phenomenology that occurs during the transition to
USR with a black hole, we use an \textit{exactly flat} potential to model the USR regime,
and do not consider the subsequent evolution beyond USR. 
In a realistic cosmology, there would need to be a further transition beyond any phase of USR to avoid eternal inflation \cite{Byrnes:2021jka}, however, such a USR-exit would
occur after the imprint of the SR/USR transition that we describe here.
A more realistic example of an ultra slow-roll potential is given in \cite{Byrnes:2018txb};
our approach could be readily extended to these more realistic potentials, but we expect 
that the spirit of our results will broadly persist.

During the SR phase of scalar evolution, the solution to \eqref{SREoMBH} is,
\begin{equation} \label{SRevolutionBH}
\varphi = \varphi_0 - \frac{W'_0}{3\gamma_{\textrm{SR}}} T_{\textrm{SR}},
\end{equation}
for some constant $\varphi_0$ that is set by the initial condition. 
From the perspective of the full equation of motion, there will be corrections to this
motion, however we can view these as QNMs that will dampen more rapidly.
The details of any generic choice of initial conditions will be washed out by the 
attraction to the SR solution \cite{Liddle:1994dx}, provided the SR period is sufficiently 
long-lived. For simplicity, we set initial conditions so that the scalar starts off exactly 
with the slow-roll velocity $|W'_0|/3\gamma_{\textrm{SR}}$, this will identify the modes 
that are excited \textit{during} the transition. While they can therefore be neglected in a 
long period of slow-roll, during USR, QNMs have a chance to dominate the solution.  

We first look at scalar evolution in the pure de Sitter cosmology, where we have 
full analytic expressions. In dS, the cosmological time, $T_c$, is also compatible with one of the USR QNMs, in the sense that $\varphi \sim \exp(-i\omega T_c)$ is one of the USR solutions found previously, with $\omega = -3Hi$.
Introducing a black hole breaks the spatial homogeneity, and the slow-roll time 
\eqref{SlowRollBHTime} now does \textbf{not} correspond to one of the dS branch USR 
QNMs (i.e.\ the modes that have purely imaginary frequencies). 
As such, through the SR-USR transition, the scalar cannot match to a single QNM, 
and instead should decay into a combination of radially-dependent modes.
We might expect with a small black hole that this effect is mild, and the decay is 
predominantly into the relevant de Sitter branch QNM. 
However, for sufficiently large black holes, we expect the additionally excited modes 
to become more significant to the overall solution during USR, which is
what we observe in our results below.

\subsection{de Sitter transitions} 
\label{TransitiondSExactSec}

In pure de Sitter spacetime, we can solve for the evolution through the SR-USR 
transition exactly, starting from slow-roll initial conditions. 
Working in FRW coordinates \eqref{FRWMetric}, the solution for the scalar is,
\begin{equation}
\varphi =
\begin{cases} 
\varphi_0 - v_\textrm{SR} (T_c-T_0)
& (T_c-T_0) < T_1 \; \textrm{ (SR)} \\
\\
\frac{v_\textrm{SR}}{3H} \left [1 - \exp \left(-3H \left(T_c-T_0-T_1\right) \right) \right] 
& (T_c-T_0) \geq T_1 \; \textrm{ (USR)}
\end{cases},
\label{dSSolCases}
\end{equation}
where $\varphi(T_0) = \varphi_0 < 0$ sets the initial value of the scalar at $T_0$, 
$\dot{\varphi}(T_0) = v_\textrm{SR} \equiv - W'_0/3 H$ is the precise SR velocity,
and $T_1 = 3H \varphi_0/W'_0$ is the time elapsed before the transition from SR to USR.

During slow-roll, the solution has $\ddot{\varphi} \equiv 0$ since we started 
with the slow-roll velocity. This transitions into the $\omega = -3Hi$ QNM 
directly, without exciting any additional QNMs. 
For $T_c \rightarrow \infty$, the scalar $\varphi$ asymptotes 
to $\varphi_{\infty} = |W'_0|/9H^2$, hence any deviations to the potential in a
realistic model should occur for $\varphi < \varphi_{\infty}$ to avoid eternal inflation.

\subsection{Transitions with a black hole}

In the case of an SR-USR transition with a black hole, we need to solve 
the problem numerically, as even the individual QNMs must be computed numerically. 
We work in static patch coordinates $(t, r)$ and numerically solve the PDE,
\begin{equation} \label{ScalarEOMstar}
\partial^2_t \psi - \partial^2_{r^*} \psi + f(r) \left(\frac{f'(r)}{r} \psi 
+ r \frac{\partial V}{\partial \varphi}\bigg|_{\varphi = \psi/r} \right) = 0,
\end{equation}
in the SdS background. Here, $\psi = r \varphi$, $f(r)$ is given in \eqref{Backgroundfr},
and the tortoise coordinate $r^*$ is given in full in \eqref{TortoiseFullExpr}. In our numerical results, we work in units with $M_p = 1$.
Since the scalar will attract to slow-roll after a sufficiently long time, we use the 
initial profile of the scalar from slow-roll given in \eqref{SRevolutionBH}, with $\varphi_0 = -1$.
(Additional details of the numerical procedure and the initial conditions are given in 
Appendix \ref{NumFitApp}.)
After solving \eqref{ScalarEOMstar} for $\psi$ numerically on some $(t, r^*)$ grid, 
we can recover $\varphi$ and use $\eqref{TortoiseFullExpr}$ to map the data onto a $(t, r)$ grid.
\begin{figure}[t]
    \centering 
        \begin{tikzpicture}[overlay]
            \draw[->, very thick, black] (-3.5,0) to[out=30, in=150] (0.2,0);
            \draw[->, very thick, black] (1.1,0) to[out=30, in=150] (4.8,0);
        \end{tikzpicture}

    \begin{minipage}{0.9\textwidth}
        \includegraphics[width=\textwidth]{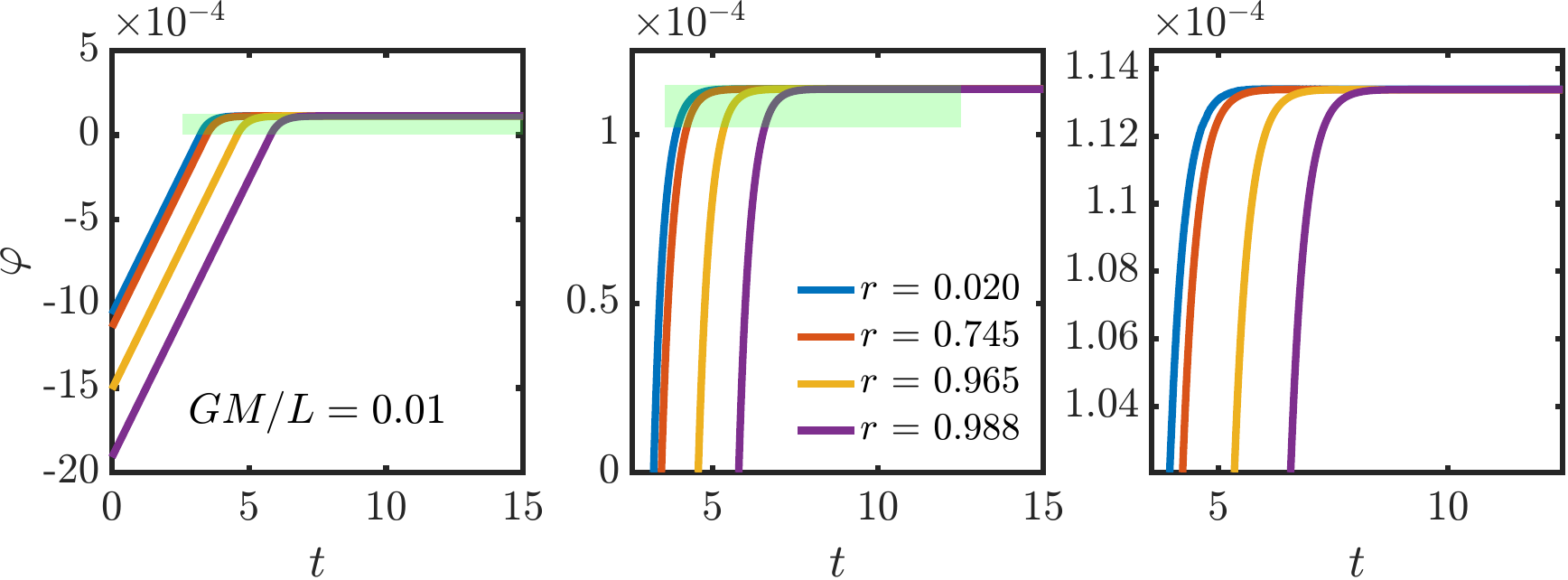}
        \vspace{-0.8em}
    \end{minipage}
    
    \begin{minipage}{0.9\textwidth}
        \includegraphics[width=\textwidth]{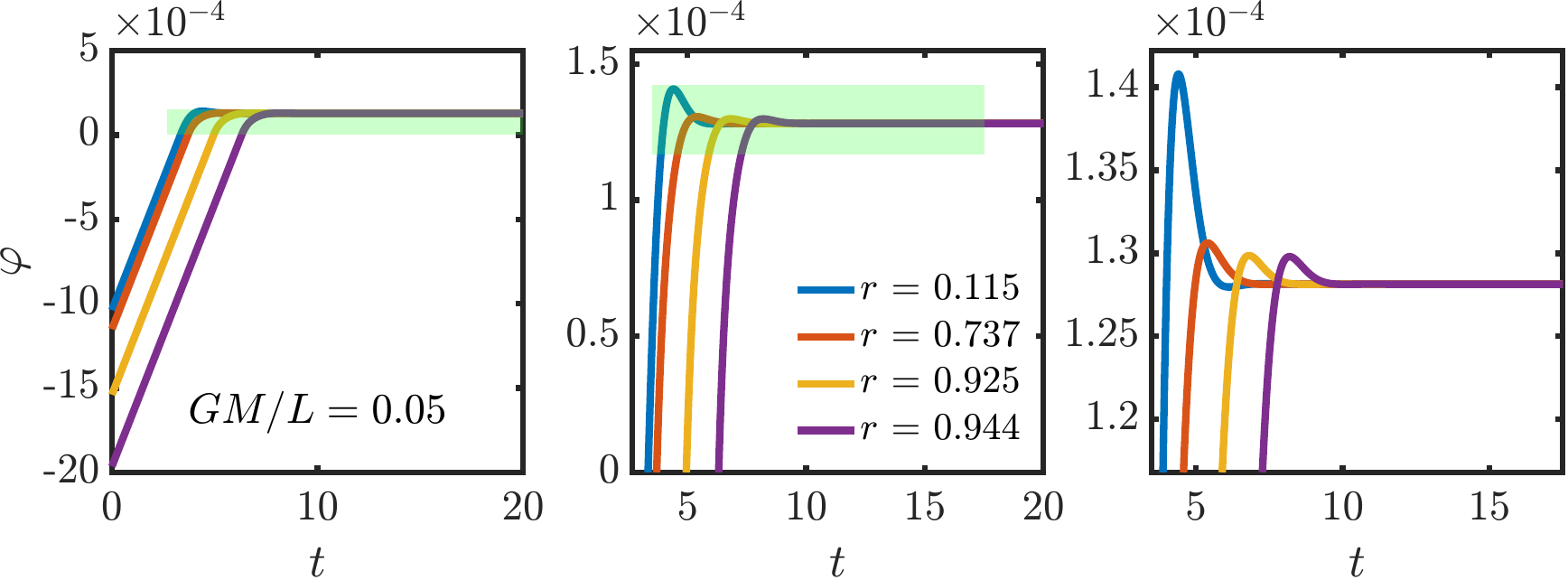}
        \vspace{-0.8em}
    \end{minipage}

    \begin{minipage}{0.9\textwidth}
        \includegraphics[width=\textwidth]{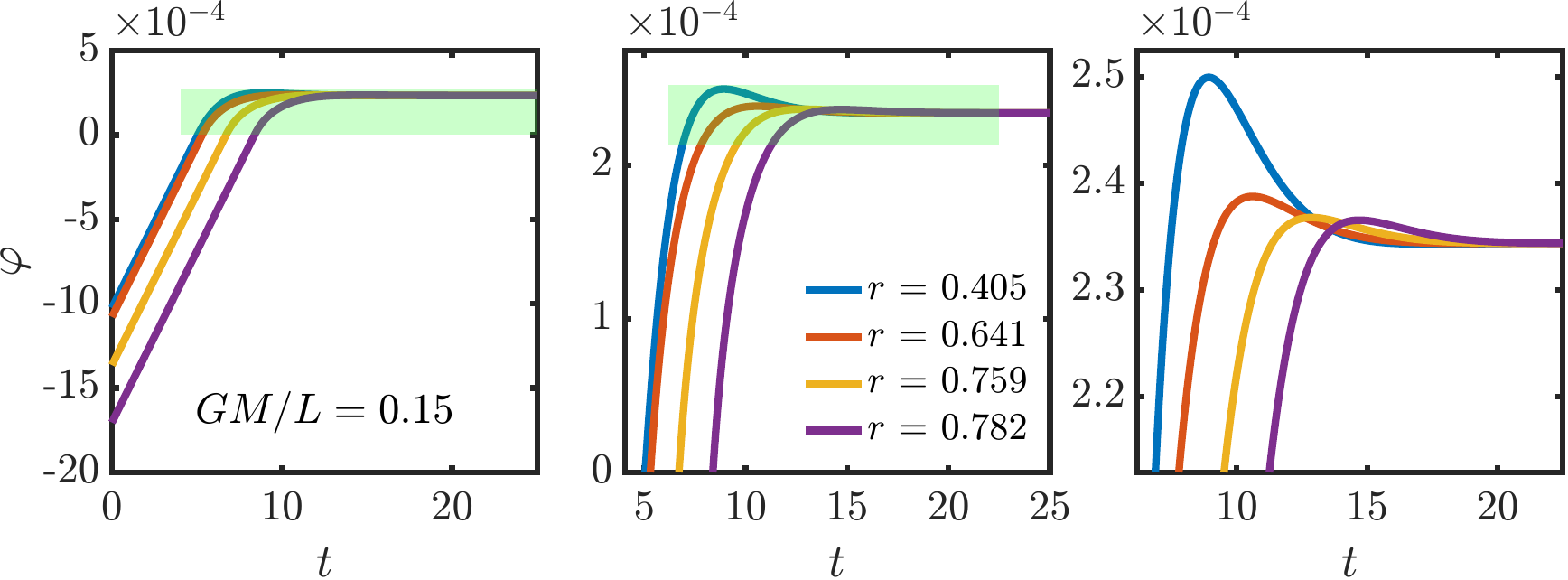}
        \vspace{-0.8em}
    \end{minipage}

    \caption{Grid of plots showing the evolution of the scalar with time $t$, for some labelled choices of characteristic radii. Each row corresponds to different $GM/L$, which are $0.01, 0.05$ and $0.15$. Each plot in the first and second columns are `zoomed' to produce the plot in the next column, for which the `zooming window' is highlighted in green. The label of $\varphi$ for the vertical axes has been omitted in some plots. The first column shows the full evolution of the scalar, where we see the scalar evolve linearly in $t$ during SR, then exponentially decay once USR commences. The second column shows the evolution from the beginning of USR onwards, and the third column shows the evolution well into USR. For small $GM/L$, during USR, the scalar experiences pure exponential decay, but for large $GM/L$, there are damped oscillations.}
    \label{fig:grid_profile}
\end{figure}
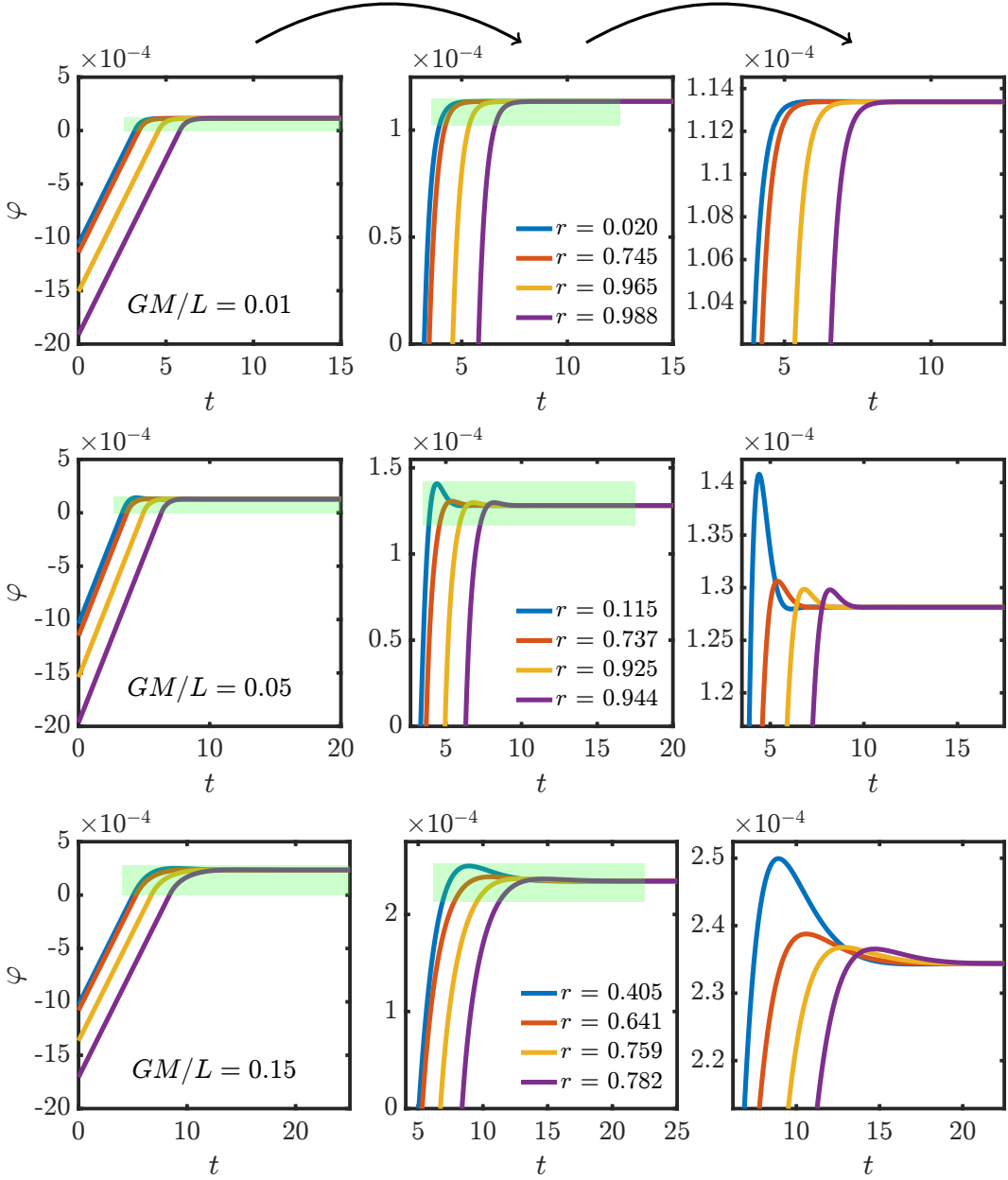

\begin{figure}[t]
    \centering

        \begin{tikzpicture}[overlay]
            \draw[->, very thick, black] (-3.8,0) to[out=30, in=150] (3.8,0);
        \end{tikzpicture}

    \begin{minipage}{0.49\textwidth}
        \includegraphics[width=\textwidth]{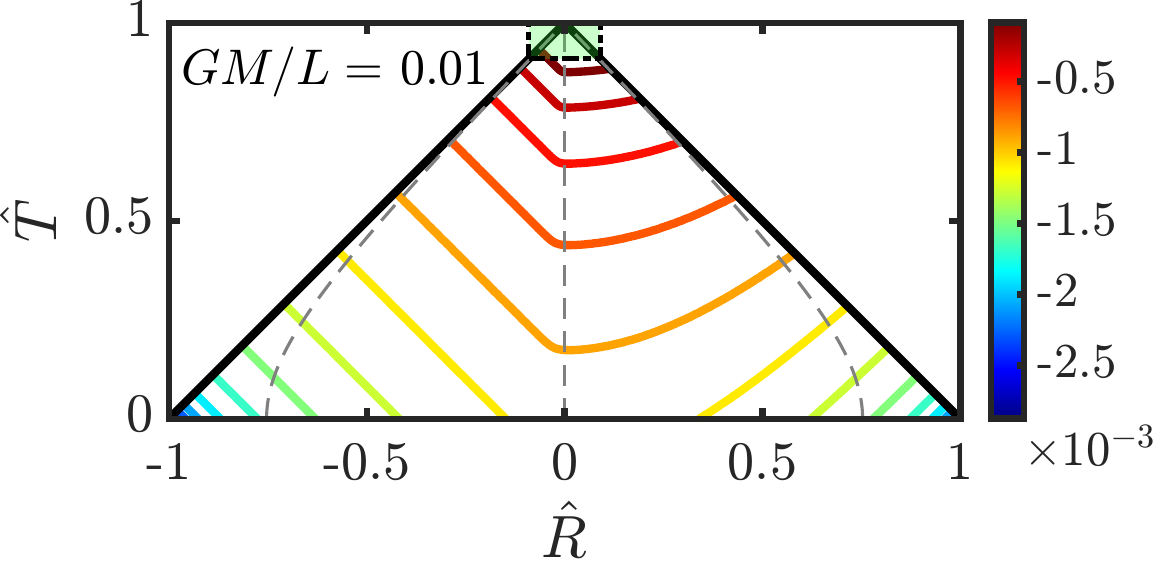}
        \vspace{-0.5em}
    \end{minipage}
        \begin{minipage}{0.49\textwidth}
        \includegraphics[width=\textwidth]{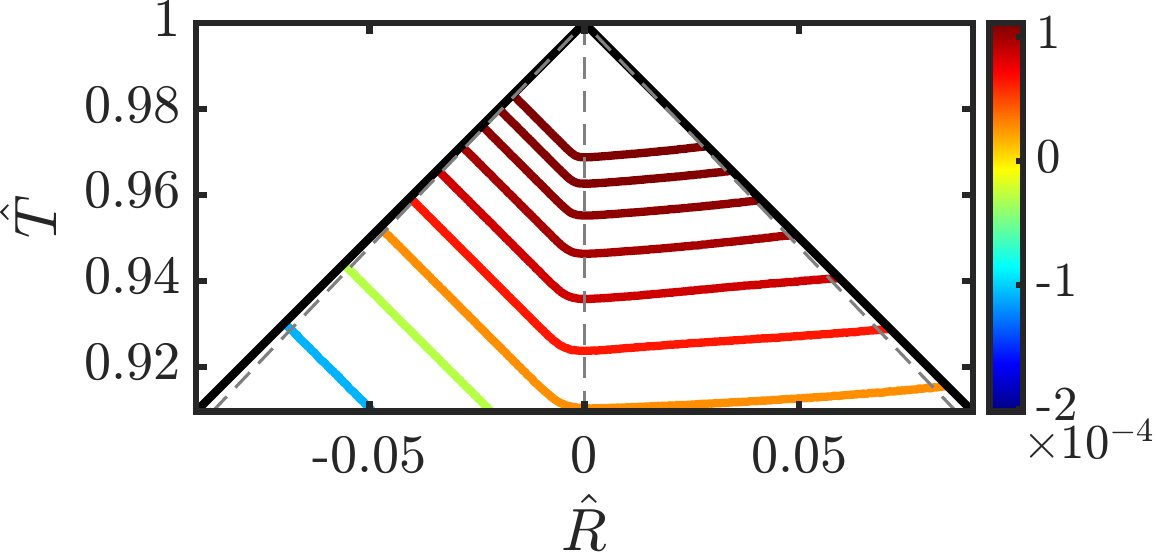}
        \vspace{-0.5em}
    \end{minipage}

    \begin{minipage}{0.49\textwidth}
        \includegraphics[width=\textwidth]{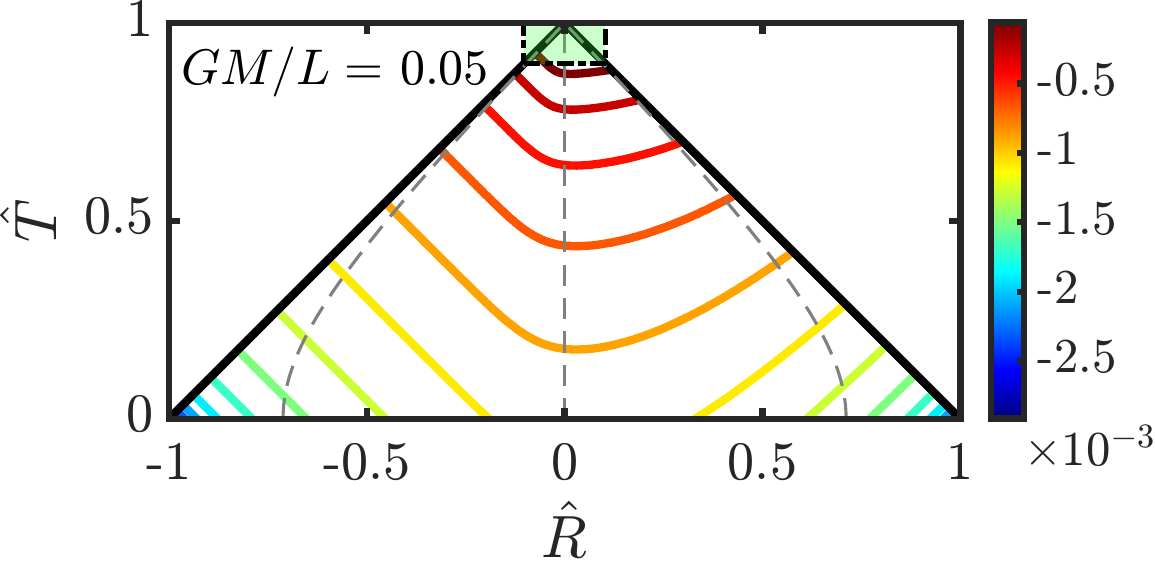}
        \vspace{-0.5em}
    \end{minipage}
        \begin{minipage}{0.49\textwidth}
        \includegraphics[width=\textwidth]{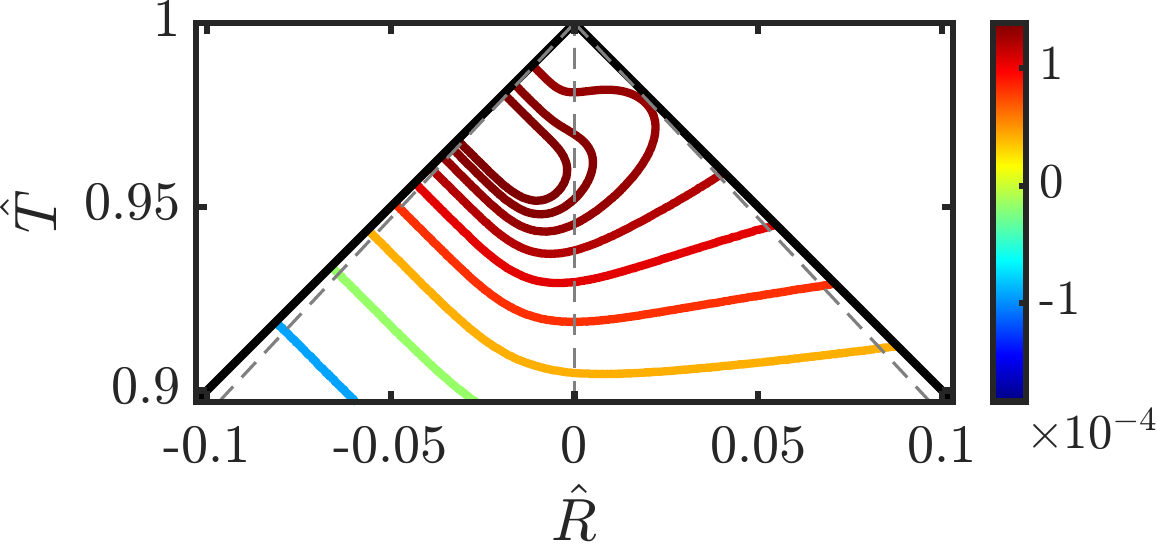}
        \vspace{-0.5em}
    \end{minipage}

    \begin{minipage}{0.49\textwidth}
        \includegraphics[width=\textwidth]{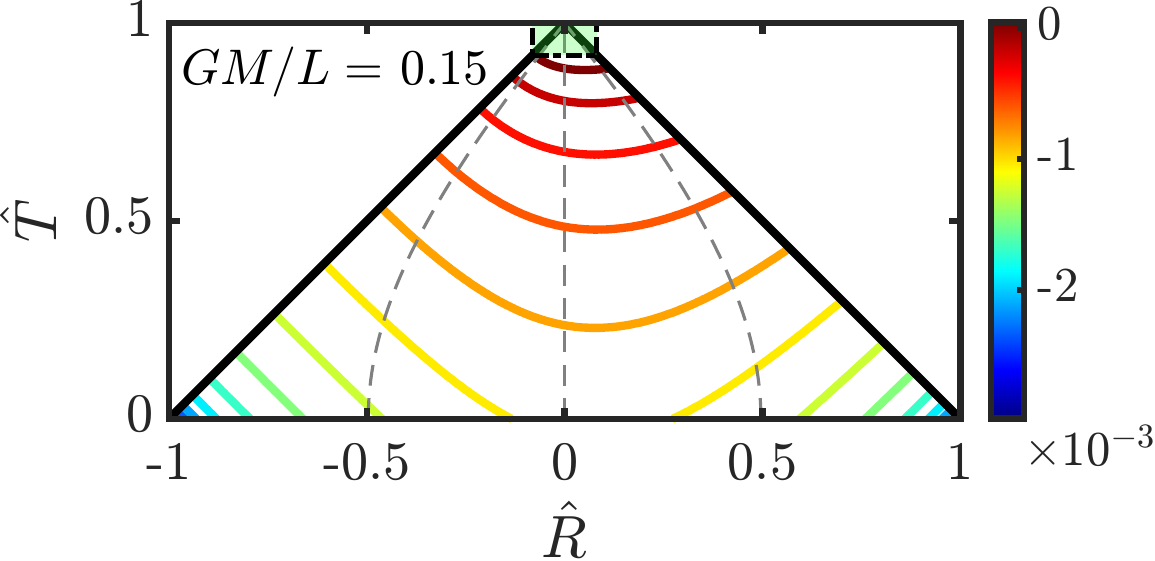}
        \vspace{-0.5em}
    \end{minipage}
    \begin{minipage}{0.49\textwidth}
        \includegraphics[width=\textwidth]{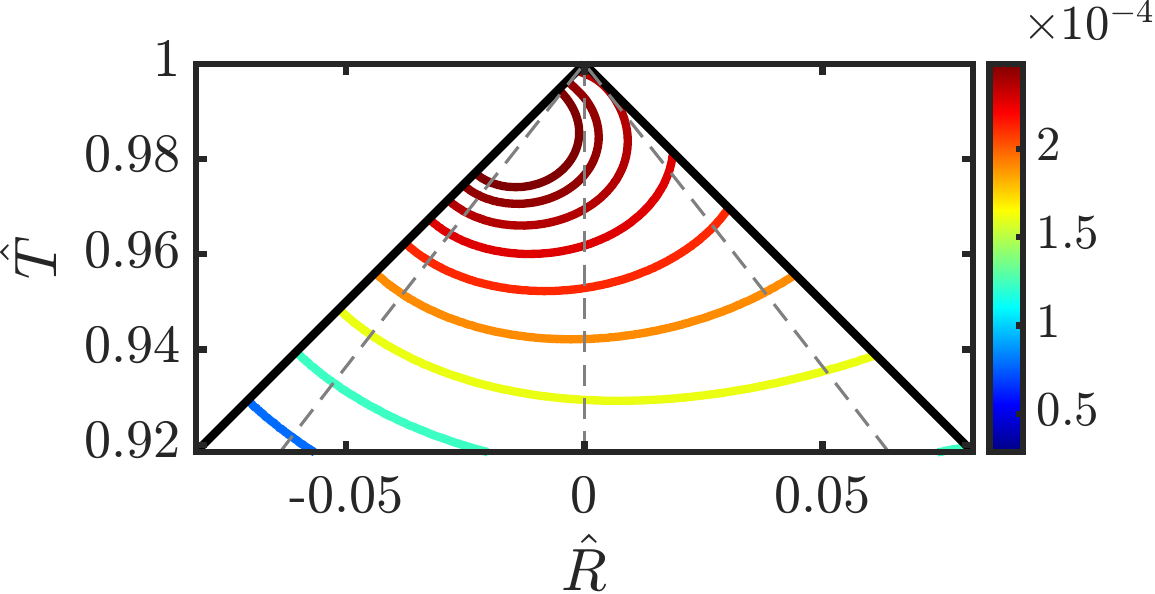}
        \vspace{-0.5em}
    \end{minipage}

    \caption{Grid of plots showing contours on Penrose diagrams of the numerical scalar solutions through the transition. Each row corresponds to different $GM/L$, which are $0.01, 0.05$ and $0.15$. The first column shows the constant contours of the scalar for the full evolution. The second column zooms into the tip of the Penrose diagram to capture some evolution during USR. The `zooming windows' relating the first and second columns are highlighted in green. These Penrose diagrams have the horizons as black diagonal lines; black hole on the left side and cosmological on the right side, and some constant radius curves are plotted in grey. For lower $GM/L$, the zoomed contours showcase pure damping, but for larger $GM/L$ we see oscillatory features manifested by the closure of the arcs. }
    \label{fig:grid_penrose}
\end{figure}

Figures \ref{fig:grid_profile} and \ref{fig:grid_penrose} give different representations of
the scalar transition for a range of black hole mass ratios $GM/L$. The solutions are 
displayed as static patch time evolutions, as well as contour plots on a Penrose diagram
(so as to illustrate how the solution might extend into the ``cosmological'' region).
In both representations, the onset of ringing behaviour is seen for $GM/L = 0.05$ and $GM/L = 0.15$. These cases correspond to the longest-lived Schwarzschild QNM being the most significant
in Fig\ \ref{ImagFreqPlot}. Note that if we start with $GM/L = 0$ (pure de Sitter), then 
the numerical solution indeed matches the analytic results above, provided we start 
with an exact slow-roll initial condition. 

Figure \ref{fig:grid_profile} shows three plots of the time evolution of the scalar at fixed radii
as indicated on the legend for three representative values $GM/L=0.01, 0.05, 0.15$. These
are taken to represent a parameter where we expect the dS QNM to be dominant, an
intermediate $GM/L$ where the Schwarzschild QNM is dominant, and a large $GM/L$ 
where we might expect the ringing effect to be most pronounced. In each case the first
plot in a row represents the full evolution for the given $GM/L$ whereas the second and third
columns zoom in successively to highlight the ringing of the scalar.
For $GM/L = 0.01$, some small oscillatory behaviour would still occur below the resolution
of our figure (which will be justified by the fit of quasi-normal modes to the signal in 
\S \ref{FittingSec}). However, since for small $GM/L$, the oscillatory Schwarzschild modes 
appear with a large negative real part (as in Figure \ref{ImagFreqPlot}), their contribution 
would quickly dampen away, leaving the remaining signal which still dampens to a constant, 
but on a slower timescale.

In Figure \ref{fig:grid_penrose} we show constant scalar contours of the same 
solutions on two Penrose diagrams; one to capture the whole time evolution, 
and one zoomed in to capture the later time behaviour in USR. 
(The construction of the Penrose diagrams is defined in Appendix \ref{Penroseapp}.) 
In the first column of plots, we show the entire evolution of the scalar on a Penrose 
diagram in each case. The horizons are shown as diagonal lines, and some constant 
radius curves are plotted as grey dashed lines. For practical reasons, these contours 
only show a reasonably early part of the evolution. Hence, most of the contours shown 
in the first column are just that of the SR evolution for each case. 
The USR evolution appears in the tip of the triangle of these plots, so in the second 
column of plots we zoom in to observe this evolution. We zoom sufficiently far to 
capture the first oscillation, if it can be numerically resolved. 
The `zooming windows' are highlighted in green in the first column of plots. 
For smaller $GM/L$, these contours do not change significantly compared to their 
SR counterparts, however, for larger $GM/L$, we see significant deviations in the 
contours, where they eventually change from being timelike to spacelike in some 
regions. The oscillatory behaviour is very clear in the last two plots, where the peak 
in the oscillation appears explicitly. 

\begin{figure}[ht]
    \centering
    \includegraphics[width=0.5\textwidth]{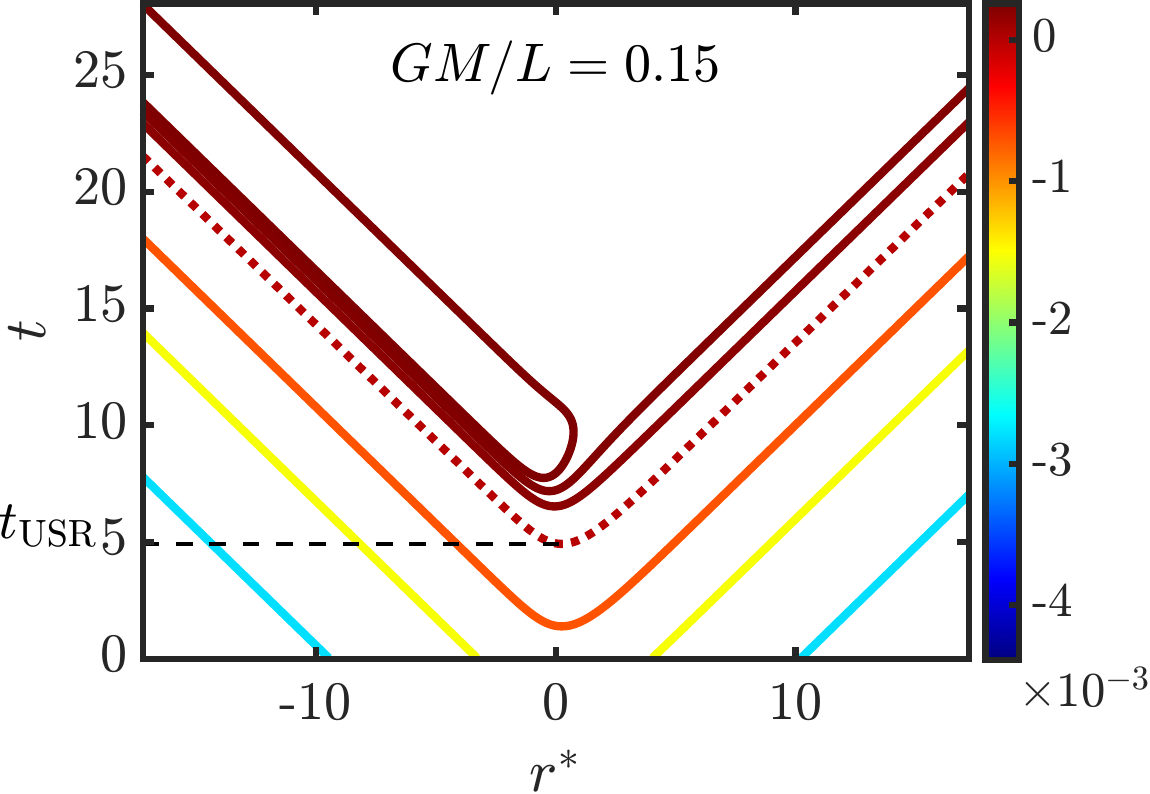}
     \caption{Some constant contours of the scalar $\varphi$ on $(r^*, t)$ axes for $GM/L = 0.15$. To emphasise the transition into USR, we have plotted the $\varphi = 0$ contour as a dotted curve. This figure highlights that for each $GM/L$, there is a radius for which USR is locally entered first. We call the corresponding static time $t_{\textrm{USR}}$, which plays a role in the fitting framework below. For $GM/L = 0.15$, $t_{\textrm{USR}} \approx 4.9$. We have only plotted a few of the contours following USR entry, which already appear clustered together, due to the exponential decay of the scalar during USR.}
     \label{t_rstar_plots}
\end{figure}

Note that, for each $GM/L$, there is a radius for which the transition into USR 
occurs earliest in time $t$. We can see this from looking at Figure \ref{t_rstar_plots}, 
which show contours of the scalar solutions on the grid $(t, r^*)$, for one particular 
case of $GM/L$ for illustration. In particular, if we consider the dotted curve in 
Figure \ref{t_rstar_plots}, which is the contour for $\varphi = 0$ (USR entry), and find 
its minimum point, this will give the value of $r^*$ for which USR occurs first locally. 
Then we can use the map between $r^*$ and $r$, given in \eqref{TortoiseFullExpr}, 
to find the radius of interest. Furthermore, there is a corresponding static time 
$t_{\textrm{USR}}$ corresponding to this first USR entry, which is significant in 
the discussion of fitting below.

\subsection{QNM fitting} \label{FittingSec}

As the scalar enters ultra slow-roll, we can ask how well the resulting data from solving the 
PDE fits to a linear combination of SdS QNMs. 
In pure de Sitter, we observed in Section \ref{TransitiondSExactSec} that the scalar decays into precisely the $\omega = -3Hi$ mode through the transition. However, with the spatial inhomogeneity of the black hole, we might expect the slow-roll scalar to split into more than one individual mode.
This is because, in contrast to what occurs for de Sitter in Section \ref{AnalyticdSQNMResultsSec}, there is no time coordinate $T$ in which the solutions are spatially homogeneous in both SR and USR.

The form of the fit that we consider for the scalar, once it reaches the USR (flat) range of the potential, is,
\begin{equation} \label{FitForm}
    \varphi = C + \textrm{Re} \, \left(\sum_n A_n \, z_n(r) \, e^{-i\omega_n (t - t_{\textrm{USR}})}\right),
\end{equation}
where $n$ is an index over the QNMs, $z_n(r)$ and $\omega_n$ are the corresponding QNM complex radial profile and frequency, and $A_n$ is a complex number. We will consider $|A_n|$ to be the amplitude of a given mode in a particular fit. $C$ is some constant, since we see (in for example Figure \ref{fig:grid_profile}) that generically the scalar comes to rest at some constant non-zero value. The constant $t_{\textrm{USR}}$ is the earliest static time $t$ that the scalar crosses into USR at $\varphi = 0$, an example of which is shown in Figure \ref{t_rstar_plots}, where $t_{\textrm{USR}} \approx 4.9$. This is included because while the simulation begins at $t = 0$, USR does not begin at any radius until later (as in Figure \ref{t_rstar_plots}). As such, for the amplitudes $|A_n|$ to be a more meaningful measure of the importance of a particular mode, we need to begin the fit closer to the start of USR\footnote{For a concrete example of the reason to include $t_{\textrm{USR}}$, consider the comparison of two fictitious modes $e^{-t}$ and $10 \, e^{-3t}$. Suppose some feature that we are modelling begins at $t = 5$. It may appear that the second mode is more significant, due to the higher amplitude in this representation, but by $t = 5$, its role is comparatively suppressed from the higher damping.}.
As noted in Section \ref{AnalyticdSQNMResultsSec}, we have fixed the overall normalisation of the $z_n(r)$ by requiring,
\begin{equation} \label{QNMprofilenormalisation}
    z_n(r) \sim e^{i \omega r^*} \quad \textrm{as } r \rightarrow r_c.
\end{equation}
In the case of the de Sitter branch of modes, the corresponding $A_n$ will be necessarily real, since both the $z_n(r)$ and the $\omega_n$ are real. However, for the Schwarzschild branch, $A_n$ can be complex in general.

In determining a measure of the accuracy of these fits, we use the \textit{root mean squared residual error}, defined by,
\begin{equation} \label{RMSREdef}
    \textrm{RMSRE} = \sqrt{\frac{1}{n} \sum \left(\frac{\varphi_{\textrm{data}} - \varphi_{\textrm{fit}}}{\varphi_{\textrm{data}}}\right)^2},
\end{equation}
where $\varphi_\mathrm{data}$ is the field in our PDE solver during USR, $\varphi_\mathrm{fit}$ is the fitting function in \eqref{FitForm} and $n$ is the number of data points on our spacetime grid used in the fit. A lower RMSRE means that the fit is more accurate.
Before fitting to \eqref{FitForm}, one needs to decide how many modes to include in the fit, as choosing too many can result in overfitting the numerical noise. We provide a detailed discussion of this point in Appendix~\ref{NumFitApp} and quote the main results here.

We find that for $GM/L < 0.012$ and $GM/L > 0.083$, the data is well fit by a single mode, which in the first (second) case corresponds to the 2nd de Sitter (1st Schwarzschild) mode.
For $0.012 \leq GM/L \leq 0.083$, we find that both modes need to be included to obtain a good fit.
We use the phrase \textit{combined mode fit} to refer to this fitting procedure; 1 mode at the edges of the $GM/L$ range, and 2 modes in the middle.

\begin{figure}[ht]
    \centering
    \begin{subfigure}{0.47\textwidth}
        \centering
        \includegraphics[width=\textwidth]{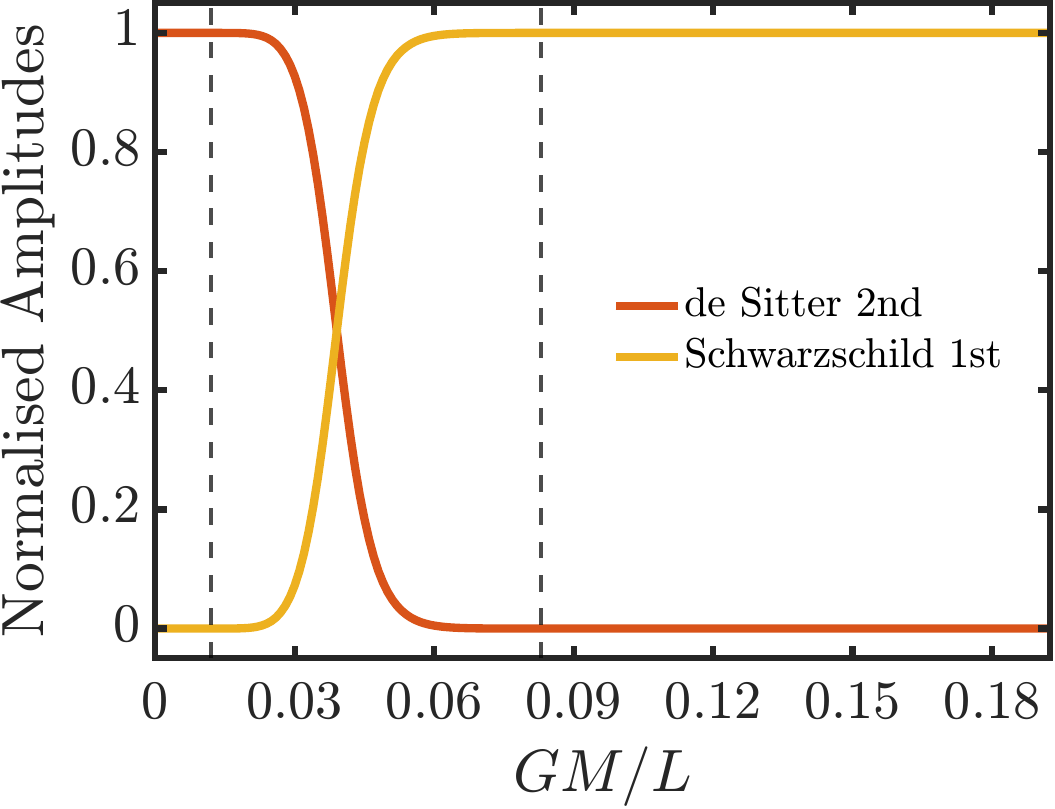}
        \caption{Combined amplitudes}
        \label{CombinedNormAmpsPlot}
    \end{subfigure}
    \hspace{0.4cm}
    \begin{subfigure}{0.47\textwidth}
        \centering
        \includegraphics[width=\textwidth]{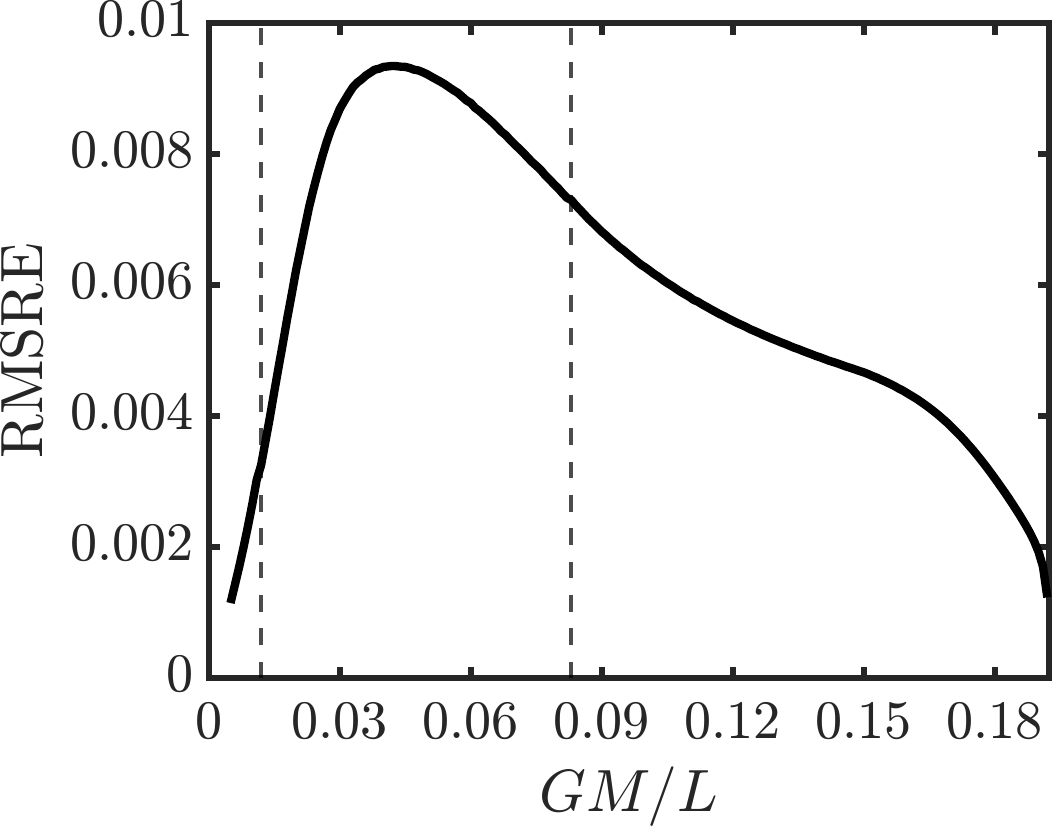}
        \caption{Combined mode fit RMSRE}
        \label{CombinedRMSREPlot}
    \end{subfigure}
    
    \caption{(a) shows a plot of normalised amplitudes for a range of $GM/L$ for the combined mode fit. Note how the amplitudes smoothly tend to $0$ or $1$ on each end, signalling the insignificance or domination of these modes, respectively. (b) shows a plot of root mean squared relative error for a range of $GM/L$ for the combined mode fit. The RMSRE is considerably lower for the combined mode fit than for the best single mode fit shown in Figure \ref{SingleRMSREPlot}. In both plots, vertical dashed lines are plotted at the transition between one and two mode fits, at $GM/L = 0.012$ and $GM/L = 0.083$.}
\end{figure}

The results of the amplitudes for combined mode fit are shown in Figure \ref{CombinedNormAmpsPlot}. In this plot, we have considered normalised amplitudes, for which we have set $\sum |A_n|^2 = 1$, where the amplitude definition is given in \eqref{FitForm}. We see from Figure \ref{CombinedNormAmpsPlot} that for small $GM/L$, the 2nd de Sitter mode strongly dominates, until around $GM/L = 0.024$, when the 1st Schwarzschild mode starts to have an appreciable effect (at the level of $|A_{\textrm{Sch}}|^2 \geq 10^{-2}$). Indeed in the proximity of $GM/L = 0.04$, these two modes compete with one another, and then cross over, with the 2nd de Sitter mode only having an appreciable effect until around $GM/L = 0.058$, above which the 1st Schwarzschild mode strongly dominates (where $|A_{\textrm{dS}}|^2 \leq 10^{-2}$). Vertical dashed lines show the cutoff values for the one/two mode fits.

The corresponding RMSRE of the combined mode fits are shown in Figure \ref{CombinedRMSREPlot}. We see that the combined mode fit has a maximum RMSRE of less than 1\%. Naturally, having additional modes in the fit does mean that there are additional degrees of freedom, so we might expect a decreased RMSRE on that basis.
We do still see an increase in RMSRE towards the intermediate range of $GM/L$ considered; this suggests that additional modes might also somewhat compete in this range, but certainly in a way that is subdominant to the 2nd de Sitter and 1st Schwarzschild modes.

\begin{figure}[t]
    \centering

        \begin{tikzpicture}[overlay]
            \draw[->, very thick, black] (-3,0) to[out=30, in=150] (4,0);
        \end{tikzpicture}

    \begin{minipage}{0.9\textwidth}
        \includegraphics[width=\textwidth]{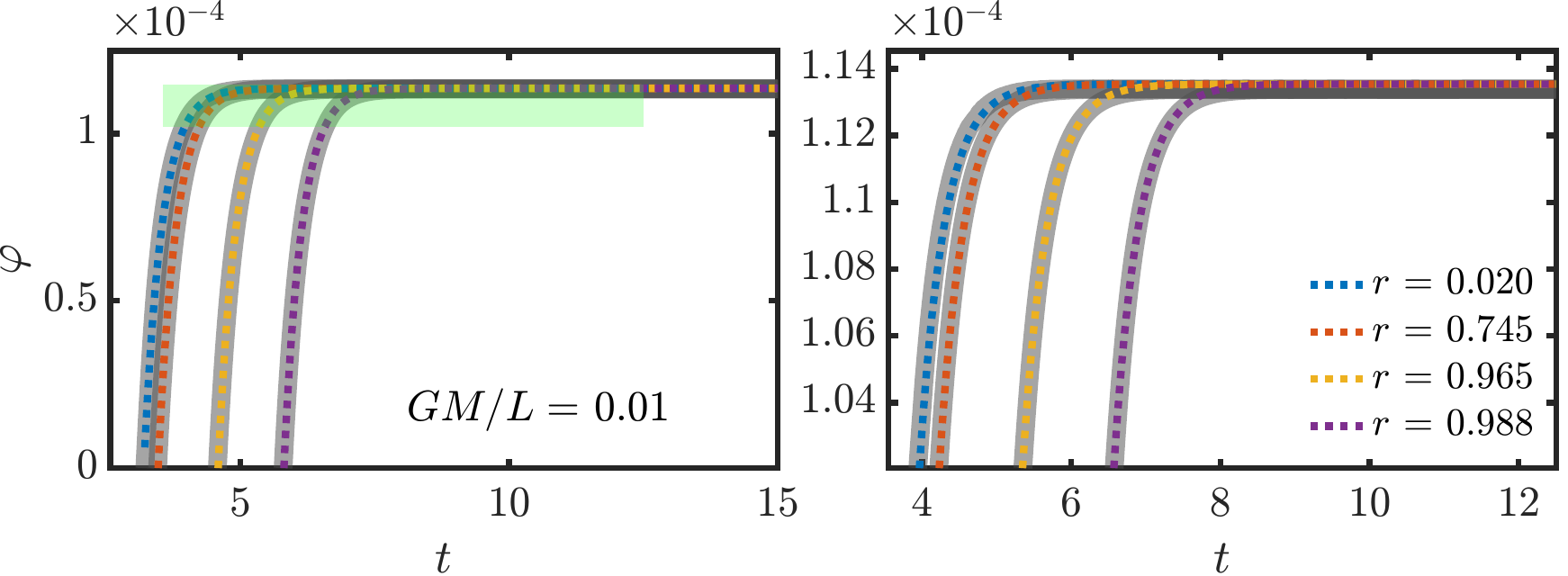}
        \vspace{-0.8em}
    \end{minipage}
    
    \begin{minipage}{0.9\textwidth}
        \includegraphics[width=\textwidth]{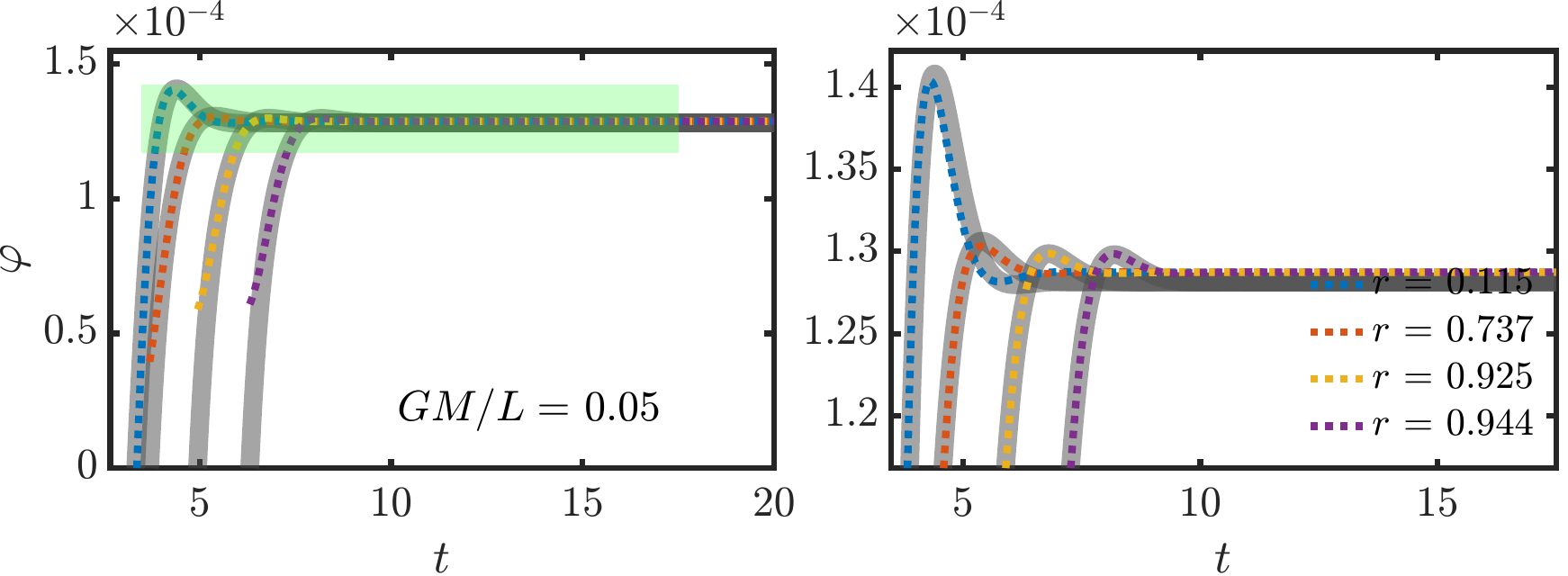}
        \vspace{-0.8em}
    \end{minipage}

    \begin{minipage}{0.9\textwidth}
        \includegraphics[width=\textwidth]{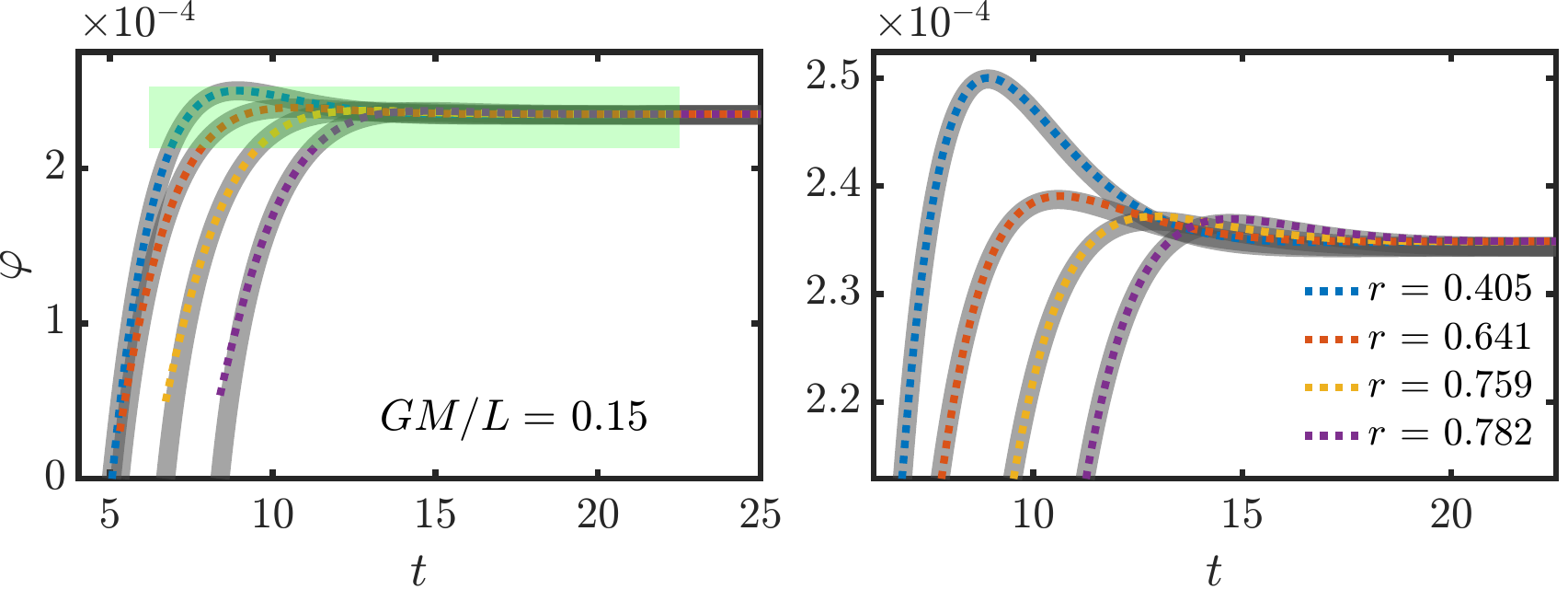}
        \vspace{-0.8em}
    \end{minipage}

    \caption{Grid of plots, showing the quality of fits of the scalar solutions to QNMs. Each row corresponds to different $GM/L$, which are $0.01, 0.05$ and $0.15$. In each plot, the grey curves are the scalar PDE solutions, and the dotted curves are the QNM combined mode fits (described in main text). The numerical scalar solutions are the same as shown in the second and third columns of Figure \ref{fig:grid_profile}; first the evolution after crossing into USR, and then an evolution within a (green) highlighted zooming window. The $\varphi$ label has been omitted in some plots. In each case, the appropriate QNMs are a good fit to the scalar solutions during USR.}
    \label{USRfits}
\end{figure}
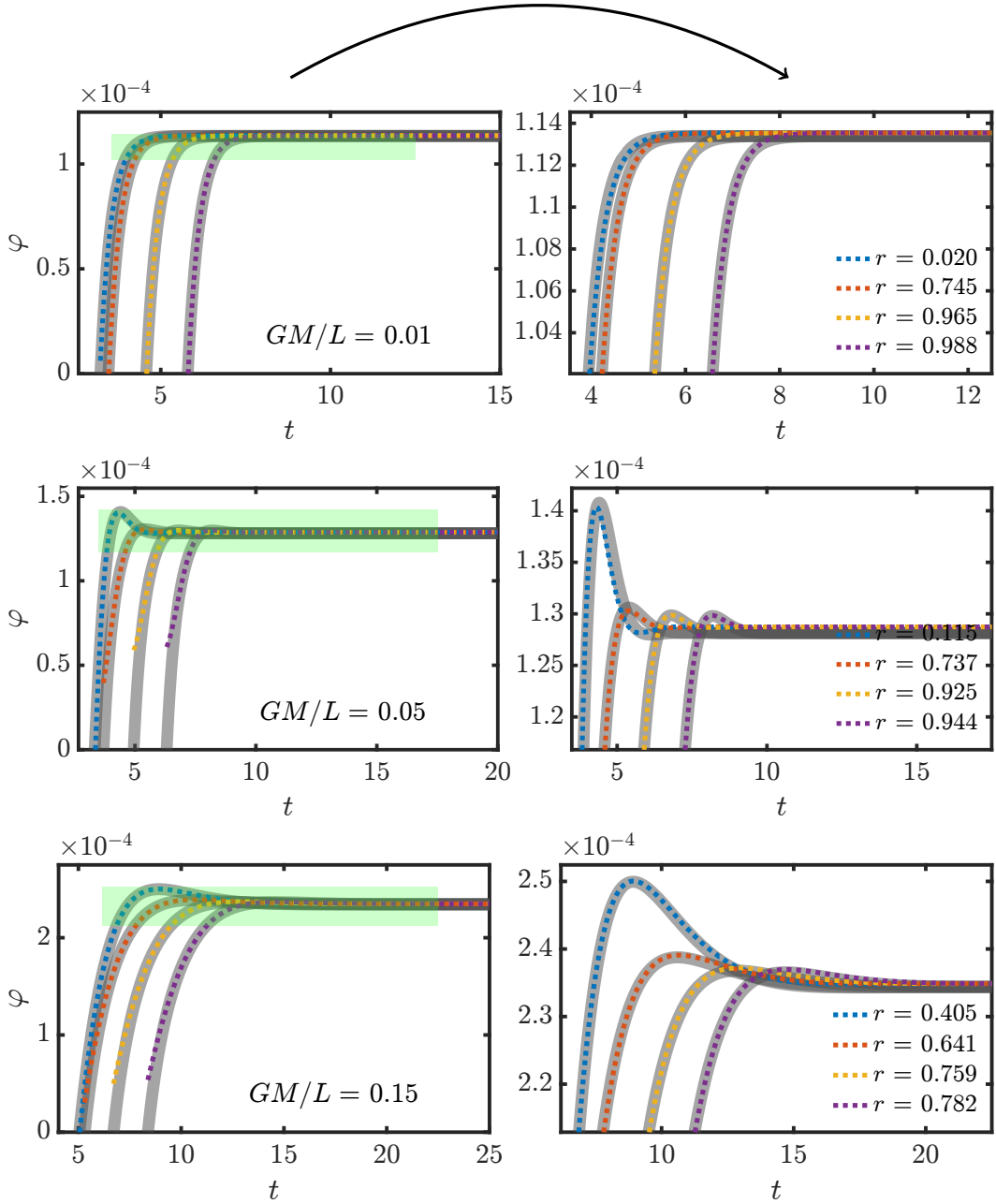

As a further confirmation of the quality of the combined mode fit, we compare the scalar evolution during USR against the fit for a range of radii. In Figure \ref{USRfits}, we show the result of fitting the data to the QNM form given in \eqref{FitForm}, for the same cases and radii given in Figure \ref{fig:grid_profile}. The first column of Figure \ref{USRfits} shows the evolution from USR entry at $\varphi = 0$, and the second column shows the same zoomed evolution, for which the `zooming windows' are highlighted in green.
Indeed we see that these linear combinations of quasi-normal modes is a good fit in each case of $GM/L$ to the scalar evolution during USR. The fit tends to be worse closer to the beginning of USR, which is expected due to residual effects of the transition, and since we do not fit precisely from the beginning of USR\footnote{There may also be power law tails at earlier times following USR entry that could also play a role \cite{Brady:1999wd}.}. The corresponding RMSRE are $2.7 \times 10^{-3}$ for $GM/L = 0.01$, $9.2 \times 10^{-3}$ for $GM/L = 0.05$ and $4.7 \times 10^{-3}$ for $GM/L = 0.15$, in line with what we observe in Figure \ref{CombinedRMSREPlot}.

Nevertheless, since we see a good fit for a range of $GM/L$, this establishes that following a transition to ultra slow-roll, the scalar $\varphi$ decays into a linear combination of quasi-normal modes. 
While we have only fitted a small number of modes to the data here (in order to avoid overfitting), our results overall suggest that many QNMs will be excited during the transition, at least with some (potentially small) amplitude, due to the inhomogeneity of the black hole. As long as the black hole is present, and the Nariai limit is not reached, this statement will be valid regardless of $GM/L$, since both branches of modes still fully appear in such cases. Of course, not all of these amplitudes will be numerically resolvable in each case, since higher modes decay faster than the dominant modes and their effect becomes comparable to the noise.

\begin{figure}[ht]
    \centering
    \begin{minipage}{0.47\textwidth}
        \includegraphics[width=\textwidth]{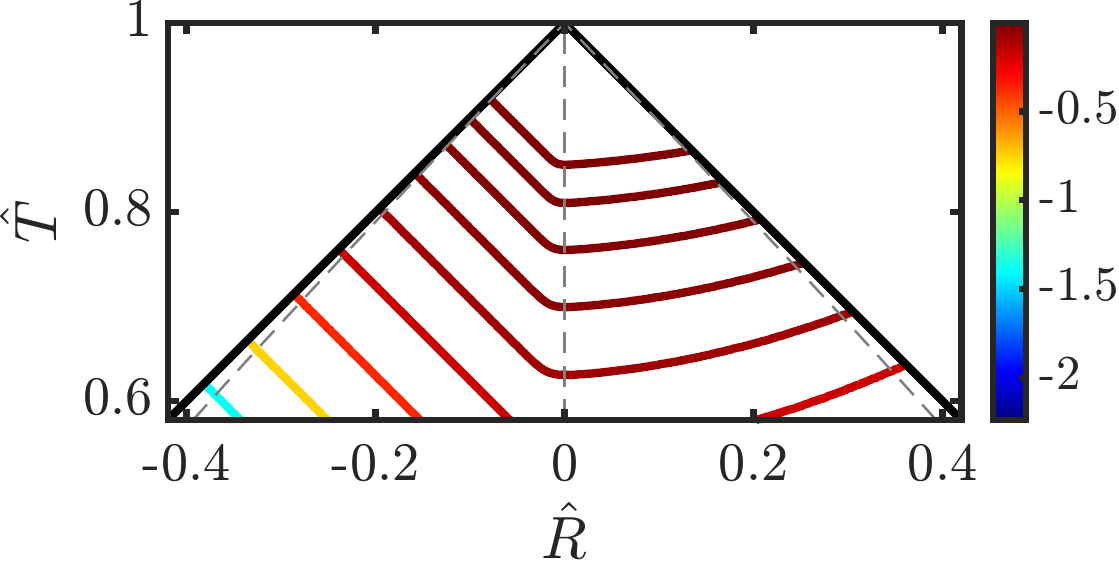}
        \caption*{$GM/L = 0.01$}
    \end{minipage}
    \hspace{0.4cm}
    \begin{minipage}{0.47\textwidth}
        \includegraphics[width=\textwidth]{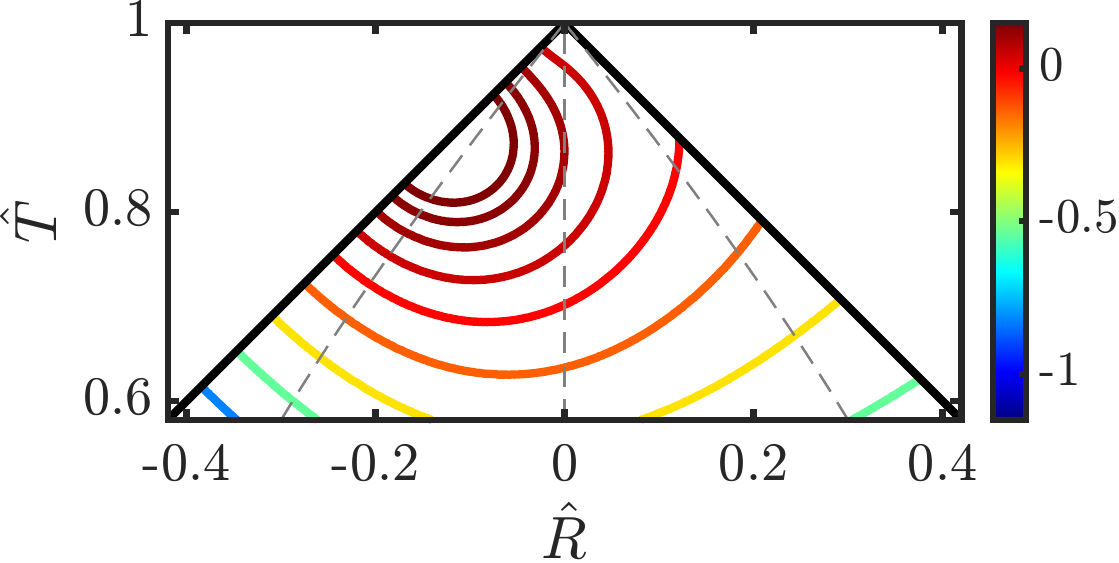}
        \caption*{$GM/L = 0.15$}
    \end{minipage}
    
    \caption{Penrose diagrams of contours of QNMs corresponding to the single mode fits, for $GM/L = 0.01$ (left) and $GM/L = 0.15$ (right). We note the similarity of these contours to the corresponding contours for the plots of the numerical solution shown in Figure \ref{fig:grid_penrose}. In particular, the features of damping and oscillatory behaviour manifest themselves in the contour shapes.}
    \label{QNMcompareplot}
\end{figure}

Now that we have fit the data to quasi-normal modes, and identified which modes dominate at each $GM/L$, we can ask whether the QNMs display the same kind of shapes in the Penrose contour plots as those that we observed in the scalar data (shown in the second column of Figure \ref{fig:grid_penrose}). We look at the two edge cases considered in Figure \ref{fig:grid_penrose}, $GM/L = 0.01$ and $GM/L = 0.15$, since we identified from the normalised amplitudes in Figure \ref{CombinedNormAmpsPlot} that a single QNM would be a good fit in those cases. 
In each case, we plot the constant contours of $ - \textrm{Re}\left(z(r) \, e^{-i\omega t}\right)$ on a Penrose diagram\footnote{We use the additional minus sign to match the overall behaviour of the scalar solutions during USR, as seen in Figure \ref{fig:grid_profile}, starting from $t = 0$ by increasing, rather than decreasing (for most of the radial grid).}, where we use the numerical mode data from QNMSpectral corresponding to the prescribed modes from the single mode fit. In particular, for $GM/L = 0.01$, we plot the 2nd de Sitter mode, and for $GM/L = 0.15$, we plot the 1st Schwarzschild mode. The results are shown in Figure \ref{QNMcompareplot}. The vertical axes and colour schemes are different to that of the corresponding plots in the second column of Figure \ref{fig:grid_penrose}, since we use a different amplitude than we find from the fit to the data. Indeed, we see similar shapes of contours as we observed in the final column of Figure \ref{fig:grid_penrose}, providing additional evidence that the fits to quasi-normal modes are valid. In particular on the right plot in Figure \ref{QNMcompareplot}, we see that one of the drawn constant radius surfaces goes through many of the near-concentric near-circular arcs, suggesting that as time varies, the QNM increases to a maximum and then decreases again (before presumably continuing to oscillate further into the future).

\subsection{Near horizon solutions}

In cosmology, one works in FRW coordinates \eqref{FRWMetric}, whereas our analysis
of the SR-USR transition has been undertaken in static patch coordinates. One might
therefore wonder about the applicability of our results to inflationary cosmology.
Since we have established a fit of the scalar field data (while it rolls in the USR range of the potential) to a linear combination 
of quasi-normal modes, here we consider individual QNM solutions, and determine 
their near (cosmological) horizon expansion to see how they extend beyond 
the cosmological horizon analytically. We then plot on Penrose diagrams to give
a visualisation of how the oscillatory behaviour continues to hold outside the horizon.
We will solve for the near cosmological horizon QNM radial profiles analytically, however 
we will input numerical values of the QNM frequencies $\omega$. 
For generic $GM/L$, we do not have analytic results for the QNM frequencies;
using our numerical results (from Section \ref{QNMSpectralNumResults}) with finite 
precision will give negligible differences to the near cosmological horizon QNMs.

To solve for the near cosmological horizon radial profile analytically, we return to the linear radial ODE \eqref{SdSRadialProfileEq}. Since at the cosmological horizon, we require $z(r) \sim e^{i\omega r^*}$ as $r \rightarrow r_c \, $, for a QNM with frequency $\omega$, we consider the substitution $z(r) = e^{i\omega r^*} Z(r)$, so that $Z(r)$ is regular at the cosmological horizon. Substituting this into \eqref{SdSRadialProfileEq} gives,
\begin{equation}
    Z'' + \left(\frac{f'(r)}{f(r)} + \frac{2}{r} + \frac{2 i \omega}{f(r)}\right)Z' + \frac{2 i \omega}{r f(r)} Z = 0.
\end{equation}
We can now find a series expansion of $Z(r)$ near $r = r_c$. Here we merely work to second order as a proof of concept of our results extending beyond the cosmological horizon. To second order, we find,
\begin{equation}
    Z(r) = Z(r_c) \left(1 + \frac{i \omega}{i \omega - |\kappa_c|} \frac{r_c - r}{r_c} + \frac{i \omega r_c f''(r_c) - 6 i \omega |\kappa_c| - 4\omega^2}{(2i\omega - 4|\kappa_c|)(2i\omega - 2|\kappa_c|)} \left(\frac{r_c - r}{r_c}\right)^2 \right) + ...,
\end{equation}
using $f'(r_c) = -2|\kappa_c|$.
To satisfy the chosen overall normalisation outlined in Section \ref{AnalyticdSQNMResultsSec}, we set $Z(r_c) = 1$.
Reintroducing the factor of $e^{i\omega r^*}$, and the time dependence, we find that the full near cosmological horizon expansion of a QNM to second order is,
\begin{equation} \label{StaticPatchQNMNH}
    \varphi \approx e^{- i \omega u} \left(1 + \frac{i \omega}{i \omega - |\kappa_c|} \frac{r_c - r}{r_c} + \frac{i \omega r_c f''(r_c) - 6 i \omega |\kappa_c| - 4\omega^2}{(2i\omega - 4|\kappa_c|)(2i\omega - 2|\kappa_c|)} \left(\frac{r_c - r}{r_c}\right)^2 \right),
\end{equation}
where $u$ is the outgoing EF coordinate defined in \eqref{EFdef}.
We can then use the cosmological Kruskal extension, as discussed in Appendix \ref{Penroseapp}, to analytically continue $r$ through the cosmological horizon. We can then use this construction to plot the near cosmological horizon expansion of a QNM on a Penrose diagram, taking the real part.

Since we work perturbatively in a near cosmological horizon expansion, we can only trust such an approximation up to some deviation from the horizon. We will consider $r$ in the range,
\begin{equation}
    \left|\frac{r_c - r}{r_c}\right| < \epsilon,
\end{equation}
where $\epsilon$ is chosen sufficiently small. We can calibrate how far the solutions can be trusted beyond the cosmological horizon by using our QNM results in the static patch. However, since the region outside the cosmological horizon is obtained by analytic continuation, this will only be a rough guide of how far we can trust the solutions beyond the horizon.

\begin{figure}[t]
    \centering
    
        \begin{tikzpicture}[overlay]
            \draw[->, very thick, black] (-5,0) to[out=30, in=150] (4,0);
        \end{tikzpicture}
        
    \begin{minipage}{0.46\textwidth}
        \includegraphics[width=\textwidth]{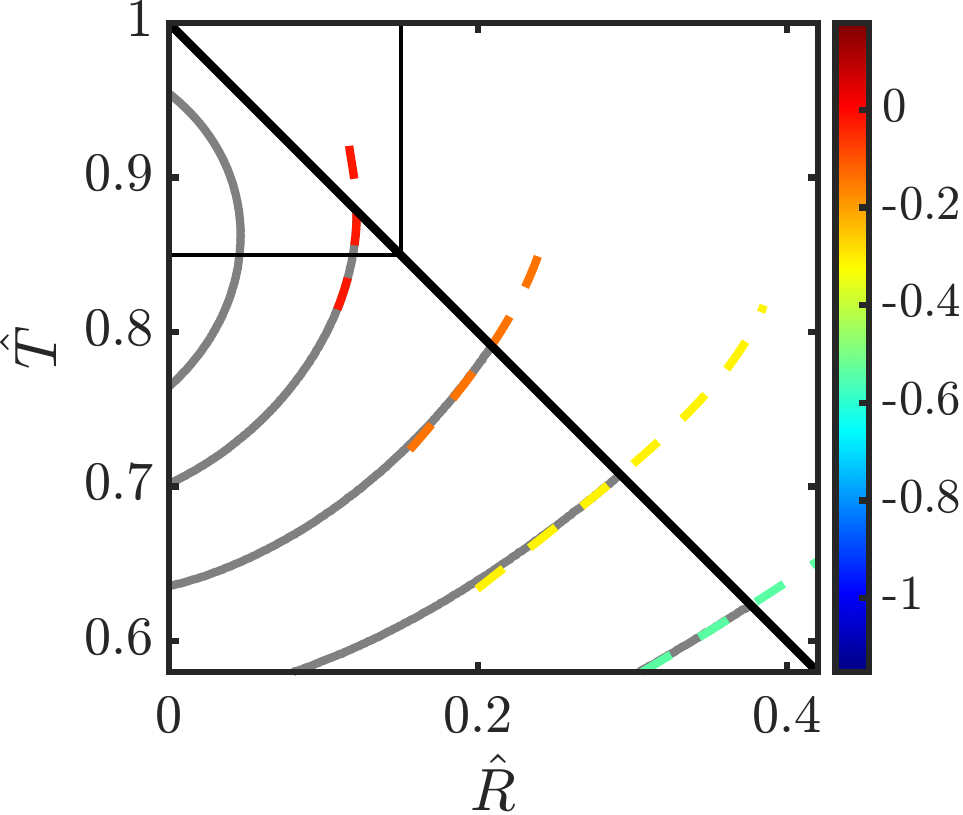}
    \end{minipage}
    \hspace{0.4cm}
    \begin{minipage}{0.49\textwidth}
        \includegraphics[width=\textwidth]{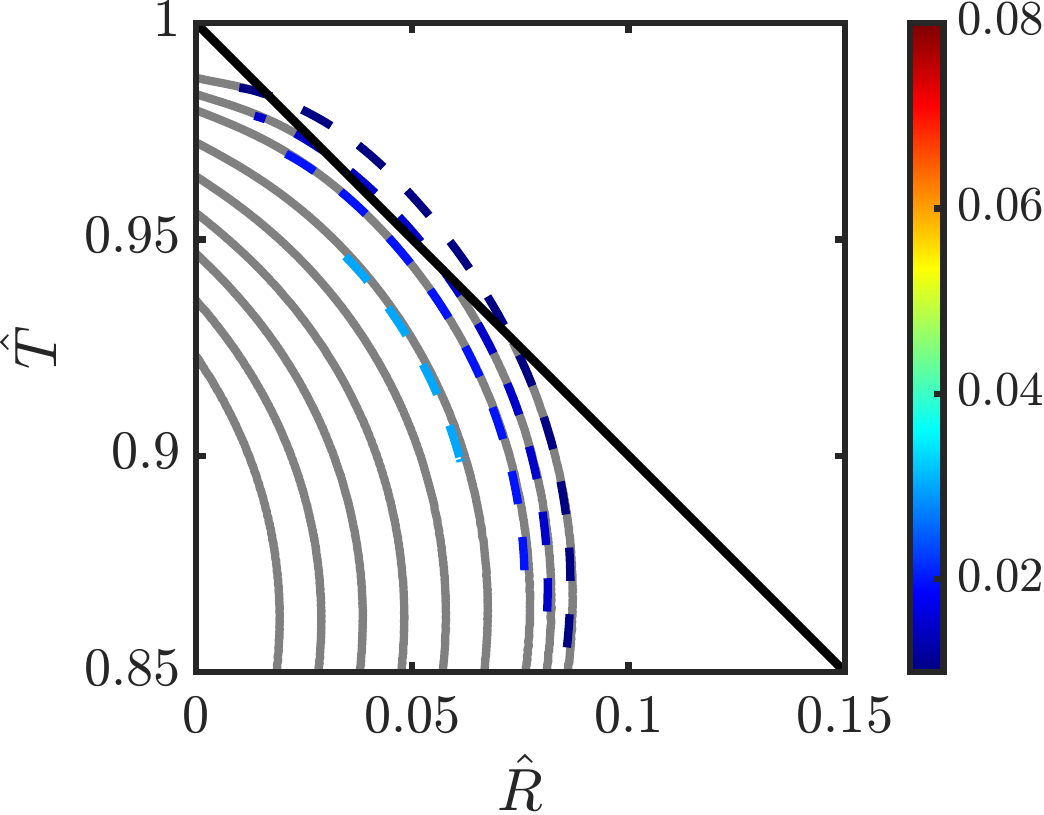}
    \end{minipage}
    
    \caption{Penrose diagrams of contours of the near cosmological horizon expansion of the dominant QNM for $GM/L = 0.15$ and $\epsilon = 0.25$. The numerical results of the QNM contour from QNMSpectral are shown as solid curves (in grey), and the near cosmological horizon expansions are shown as dotted curves (in colour). The left plot compares the near-horizon expansion to the numerical results of single mode fits shown in Figure \ref{QNMcompareplot}, which shows good agreement in their region of mutual overlap. The right plot shows a further zooming in (shown with a thin rectangle on the left plot) to the oscillatory feature, showing that it persists outside the cosmological horizon.}
    \label{QNMNHplot}
\end{figure}

We intend to verify that if oscillatory features occur within the static patch, then they can persist beyond the cosmological horizon. From Figure \ref{QNMcompareplot}, we see oscillations at the level of the Penrose diagram for $GM/L = 0.15$, which manifest in this case as near-concentric near-circular arcs.
We stress that this oscillatory behaviour should occur for all cases where a black hole is present, but that we will not be able to numerically resolve it unless the black hole is sufficiently large, and as observed in Section \ref{FittingSec}, the oscillatory modes do not dominate unless $GM/L \gtrsim 0.058$.
As a representative case, we can plot the near cosmological horizon expansion of the dominant QNM for $GM/L = 0.15$, using \eqref{StaticPatchQNMNH} and the Kruskal extension map given in \eqref{UVrmap}.
The results are shown in Figure \ref{QNMNHplot}, for $\epsilon = 0.25$. The solid grey curves are the numerical results of the QNM contours from QNMSpectral, which are valid in the static patch. The dotted curves are the near cosmological horizon extension. We only show the results for $\hat{R} \geq 0$; the region of the diagram for which the cosmological horizon is depicted. Further details on the Penrose construction are given in Appendix \ref{Penroseapp}. 
In the left plot of Figure \ref{QNMNHplot}, the colours and contour levels are chosen to match the colours of Figure \ref{QNMcompareplot}. We see that the curves overlap well in their region of mutual validity (the static patch), suggesting that $\epsilon = 0.25$ is sufficiently small for this case. In order to consistently consider a larger $\epsilon$, we would likely require a higher order expansion.
In the right plot, we zoom in on a region where the contours push just outside the horizon, and come back within. We choose a different set of contour levels to focus on this behaviour.
Nonetheless, we indeed see that the near-concentric near-circular arcs are completed in the near cosmological horizon expansion for $GM/L = 0.15$. This suggests that the oscillatory feature of USR with a black hole persists outside the cosmological horizon.

\section{Discussion}

Ultra slow-roll inflation offers an exciting prospect for enhancement of the peak of the
primordial power spectrum, which could lead to additional PBH formation, the bulk of which is 
sourced by the SR/USR transition.
In this work, we have considered the impact of a black hole on this transition. 
An innate seed black hole, present at the time of an SR-USR transition, can alter the dominant decay channel into USR from a purely damped mode (for a small seed black hole), to a damped ringing mode (large black hole). These alternative channels for the decay into USR modify the nature of the cosmological background. Cosmological perturbation theory is performed around a homogeneous FRW background, with the addition of a black hole, however, the background is now non-homogeneous in space. 
Furthermore, we have shown that for large black holes the dominant behaviour in the scalar arises from ringing (or QNM) modes, therefore there will now be oscillatory behaviour \textit{in the background}.   
We have also demonstrated that QNM features, such as ringing, extend beyond the cosmological horizon in static coordinates.
It is therefore reasonable to expect that the black hole produces ripples in the asymptotic cosmological background.
This will then feed through to further excitations of the cosmological perturbations, which would likely give modified enhancement, affecting the production of primordial black holes.
To quantify this effect we would first need to analytically continue from static coordinates beyond the cosmological horizon to have an explicit form for the full cosmological metric, which would now be not only time dependent but oscillatory and with spatial modulation, a further complication being that this background is only known numerically. One would then need to perform perturbation theory on this more complicated background, and it is likely that the usual decomposition into scalar, vector and tensor modes would be modified. 

We conclude by highlighting other aspects of our study that offer opportunities for future directions and improvement on our approximations.
As noted in Section \ref{TransitionSec}, the potential we have considered is quite simplistic. The details of a more realistic potential featuring an SR-USR transition could modify the conclusions of the dominating modes in Section \ref{FittingSec}, but we broadly expect that these conclusions will be fairly robust, while only mildly modifying where the crossover of competing modes can occur.
Furthermore, our framework only applies within the limitations of the test field approximation, and alternative models for the potential could violate this. We have worked within these limitations here to provide an illustration of the phenomenology we observe in a simple scenario where we have more control. One could derive the coupled equations for the scalar and metric evolution, and proceed to solve numerically (but this may be computationally expensive). If the metric were allowed to evolve, we may observe the decay into QNMs of the resulting dynamical spacetime. Dynamical ringdown has been of recent interest, and has been explored in \cite{Redondo-Yuste:2023ipg, Zhu:2024dyl}.
Another remark is that we only consider only a single black hole within the spacetime; if there are any black holes as inflation begins, then there are likely to be many black holes. We work within this limitation as it offers analytic control of the background metric. There are relatively few analytic examples of multi-black hole metrics; a cosmological example is \cite{Kastor:1992nn}.

In this work, for simplicity we have only considered spherical, $SO(3)$ symmetry. However, rotating black holes are more astrophysically relevant, and it is possible that an innate black hole could have angular momentum. We conclude by outlining how to treat slow-roll with rotation.

The Kerr-de Sitter metric describes such a solution,
\begin{equation} \label{kds1}
\begin{split}
g_{\mu\nu}dx^\mu dx^\nu = & \ -\frac{\Delta_r}{\rho^2}\left(dt-\frac{a\sin^2\theta}{\Xi}d\phi\right)^2 + \frac{\Delta_\theta\sin^2\theta}{\rho^2}\left(adt-\frac{r^2+a^2}{\Xi}d\phi\right)^2 \\
& \qquad \qquad + \frac{\rho^2}{\Delta_r}dr^2 + \frac{\rho^2}{\Delta_\theta}d\theta^2,
\end{split}
\end{equation}
where $a$ represents the rotation parameter, and we have,
\begin{equation} \label{defs1}
\begin{split}
\Delta_r & \ = (r^2+a^2)\left(1-\frac{r^2}{\ell^2}\right) - 2Gmr, \qquad \Delta_\theta = 1+\frac{a^2}{\ell^2}\cos^2\theta, \\
\rho^2 & \ = r^2 + a^2\cos^2\theta, \qquad \Xi = 1 + \frac{a^2}{\ell^2}.
\end{split}
\end{equation}
It is straightforward to generalise the ``slow-roll with a black hole'' scenario discussed in Section \ref{SRwithaBHSec}, to allow for an angular dependence of the scalar,
\begin{equation} \label{T_trans}
\begin{split}
T_{\textrm{SR}} = & \ t + \int dr\, \frac{r^2+a^2}{\Delta_r} \left(-\gamma r + \frac{\beta}{r^2+a^2} \right) - \int d\theta\, \frac{\gamma a^2 \sin\theta\cos\theta}{\Delta_\theta}, \\
\chi_{\textrm{SR}} = & \ \phi + \int dr\,\frac{a\Xi}{\Delta_r} \left(-\gamma r + \frac{\beta}{r^2+a^2} \right),
\end{split}
\end{equation}
where the constants $\beta$ and $\gamma$ are,
\begin{equation} \label{3gamma}
\gamma = \frac{r_c^2+r_b^2+2a^2}{r_c(r_c^2+a^2)-r_b(r_b^2+a^2)}, \qquad \beta = \frac{(r_c^2+a^2)(r_b^2+a^2)(r_c+r_b)}{r_c(r_c^2+a^2)-r_b(r_b^2+a^2)},
\end{equation}
in terms of the analogous horizon radii $r_b$, $r_c$.
Interestingly, $\gamma$ can be expressed as,
\begin{equation} \label{AreaVol}
3\gamma = \frac{\mathcal{A}_c+\mathcal{A}_b}{V_c-V_b},
\end{equation}
which is the sum of the horizon areas divided by the difference in thermodynamic volumes.
This agrees with the findings of \cite{Gregory:2018ghc} for the non-rotating case.

With a rotating black hole, it would be interesting to determine the hierarchy of dominant modes in the decay following the SR-USR transition. For slowly-rotating black holes, we would expect to see broadly similar results. However, some interesting phenomenology could emerge for larger rotation.

\acknowledgments

The authors thank Andrew Gow, Betti Hartmann, Ted Jacobson and David Wands for useful discussions. LC is supported by King’s College London through an NMES funded studentship. RG is supported in part by the STFC grant (ST/X000753/1) and in part by the Perimeter Institute for Theoretical Physics. SP is supported by the STFC, as part of the UKRI Quantum Technologies for Fundamental Physics program (grants: ST/T005858/1 and ST/T006900/1). LC and SP are grateful for the hospitality of Perimeter Institute where part of this work was carried out. Research at Perimeter Institute is supported in part by the Government of Canada through the Department of Innovation, Science and Economic Development and by the Province of Ontario through the Ministry of Colleges and Universities. This research was also supported in part by the Munich Institute for Astro-, Particle and BioPhysics (MIAPbP), which is funded by the Deutsche Forschungsgemeinschaft (DFG, German Research Foundation) under Germany´s Excellence Strategy – EXC-2094 – 390783311 (RG).

\appendix

\section{Numerical implementation and fitting details} \label{NumFitApp}

In this appendix, we describe the numerical integration of \eqref{ScalarEOMstar}.
First, we discretise the $(t,r^*)$ grid, using $10^4$ uniformly spaced grid points in $r^*$ spanning from $r^*_-$ to $r^*_+$.
We treat all functions of $r$ in \eqref{ScalarEOMstar} as functions of $r^*$.
Since the inversion of $r^*=r^*(r)$, given in \eqref{TortoiseFullExpr}, does not exist in closed form, we instead compute $r^*$ on a sample $r$ grid, then linearly interpolate to find the values of $r$ on our chosen $r^*$ values. Near the horizons, we invert $r^*(r)$ to leading order in $(r-r_{b,c})$ to extend our grid closer to the horizons.
The second spatial derivative is implemented using a three point centred finite difference stencil.
At $r^*_-$ ($r^*_+$) we assume an in-going (out-going) boundary condition i.e. $\psi(r^*\to r^*_\pm)\sim f(t\mp r^*)$, to ensure regularity.
This form of the solution is exact in the limit $r^*_\pm\to\pm\infty$.
In our code, we take $|r^*_{\pm}|$ to be sufficiently large, such that the edges of our domain approximate the spatial locations of the black hole and cosmological horizons.
We use the values $r^*_{\pm} = \mp r^*_{\textrm{max}}$, where $r^*_{\textrm{max}}$ is chosen sufficiently large to ensure that the edges of the resulting grid of $r$ values are exponentially close to the horizons, but small enough so that there are enough gridpoints spread throughout the spatial bulk. Representative example pairs of satisfactory $(GM/L, r^*_{\textrm{max}})$ are $(0.01, 12.4), (0.15, 17.3)$ and $(0.19, 76.8)$, with a drastic increase in $r^*_{\textrm{max}}$ towards the Nariai limit since $\kappa_{b,c} \rightarrow 0$ there.
The boundary conditions are encoded into the second spatial derivative using the method described in Appendix A of \cite{solidoro2024quasinormal}.

For our initial condition, we take $\varphi$ to be the SR solution \eqref{SRevolutionBH} (with $\varphi_0 = -1)$, which is justified since the SR is an attractor on the sloping part of the potential, i.e. any deviations from this solution will dissipate away exponentially.
There is a subtlety involved in this construction which we briefly elaborate.
If the field is exactly in its slow-roll configuration then $\varphi(r^*\to \pm\infty)\to -\infty$.
However, one does not expect the potential in Figure  \ref{PotentialExamplePlot} to grow linearly as $\varphi \rightarrow -\infty$.
This is really an artefact of the static coordinates, since an observer whose proper time is measured by $t$ will never see the field cross the horizon.
Hence, the configuration of the field in the far past will be imprinted in the near horizon region.
This means that if the potential does something beside grow linearly for $\varphi\to-\infty$ (for example if there is another plateau in this range of the potential \cite{Baumann:2009ds}) then expressing \eqref{SRevolutionBH} in static coordinates will not give the correct form of the field near the horizon.
To circumnavigate this subtlety, we can assume that any discrepancies from the true field configuration lie outside our range of $r^*$, such that using \eqref{SRevolutionBH} in the simulated domain is a good approximation of the true solution in that region.
This construction has the benefit that the details of $\varphi$ in the far past do not affect the result, since they are so close to the horizons that they effectively pass through without reflecting back into the static patch.

Using this SR initial condition, we integrate \eqref{ScalarEOMstar} in $t$ using a fourth-order Runge-Kutta algorithm. We proceed in timesteps of $\Delta t = 0.5 \, \Delta r^*$, to ensure that the CFL-type condition is satisfied \cite{langtangen2003computational}. We integrate for sufficiently many timesteps to ensure that the vast majority of the radial grid has successfully entered USR by the end of the simulation. For the range of $GM/L$ considered in the main text, $2 \times 10^4$ timesteps is sufficient.
We ensure convergence by checking that our results are unchanged when we increase the spatial resolution (automatically increasing our temporal resolution in turn), and the values of $r^*_{\pm}$. We checked this for many representative cases, and found insignificant differences in the results compared to the parameters outlined in this appendix.

We now also discuss additional details of the quasi-normal mode fits in Section \ref{FittingSec}. In some works, authors perform QNM fits by fixing a radius $r$, and fitting the data for $\varphi$ on that spatial slice. However, given that we have the radial profiles from the QNMSpectral output, we will include them for increased accuracy \cite{Zhu:2023mzv}, using multiple radii for the fit. Also, since we want to compare the results of the fits over a range of $GM/L$, the inclusion of the radial profiles in the fit will give better consistency in such a comparison.

As such, for fixed $GM/L$, after solving the PDE numerically on a grid of $(t, r)$ for the scalar $\varphi$, to perform a fit we can:
\begin{itemize}
    \item Identify $t_{\textrm{USR}}$ by finding the radius for which the scalar crosses $\varphi = 0$ earliest (with respect to static time $t$), as we see in Figure \ref{t_rstar_plots}.
    \item Choose a collection of quasi-normal modes to include in the fit. This includes their fixed frequencies $\omega_n$, and radial profiles $z_n(r)$.
    \item Pick a range of radii $r$ from the grid to include in the fit.
    \item For each radius $r$, identify the range of timesteps to include in the fit. Since we aim to only fit the signal during USR, we certainly should pick a starting time for which $\varphi > 0$. Including data from \textit{precisely} $\varphi = 0$ onwards can give a significantly worse fit, since residual effects from the transition (such as spatial gradients from neighbouring radii which have not entered USR yet) briefly disrupt USR from occurring. Instead we choose to wait to start the fit until the scalar profile (for that fixed radius) reaches 50\% of its maximum eventual height (the constant value of that the scalar tends to at that radius, see Figure \ref{fig:grid_profile}). We then include some prescribed number of timesteps from that point onwards. 
    \item For each pair $(t, r)$ identified to be included in the fit, collect the $\varphi$ computed on that gridpoint by the PDE solver. The collection of all selected data points are then fit against \eqref{FitForm}.
    \item Use a fitting procedure to identify the best-fit complex numbers $C, \{A_n\}$.
\end{itemize}

We now outline the rationale for the combined mode fit discussed in Section \ref{FittingSec}. While we could consider any number of QNMs in the fit, if we wish to make quantitative conclusions about the importance of each mode from their amplitude, we need to be careful of overfitting.
Many authors \cite{Giesler:2019uxc, Zhu:2023mzv, Baibhav:2023clw} caution against overfitting quasi-normal modes to signals. There is some numerical noise in our signal from discretising the PDE for the scalar evolution. Due to the exponential decay in time of the QNMs, one needs to be careful not to overfit insignificant modes to the signal, as they may appear with a large unphysical amplitude, while only fitting some of the noise. 
As such, we will be conservative and consider fits involving fewer QNMs, as this will allow us to isolate the dominant behaviour. Indeed many modes are likely to be excited by the transition (when a black hole is involved), so many modes will play some (however small) role; we do not report on these other amplitudes here in order for to get meaningful results on the dominant behaviour.

In line with this conservative approach, we start by performing a fit for the data to a \textit{single} QNM. From pure de Sitter, where the SR and USR solutions are written in terms of a single (cosmological) time $T_c$, we expect the corresponding $\omega = -3Hi$ mode to dominate the signal following entry into USR. Since there are an infinite number of QNMs, and broadly we expect QNMs to contribute less to a fit if they are less long-lived, we perform a single mode fit only for the first two modes from each branch. 

\begin{figure}[ht]
\centering
\includegraphics[width=0.66\textwidth]{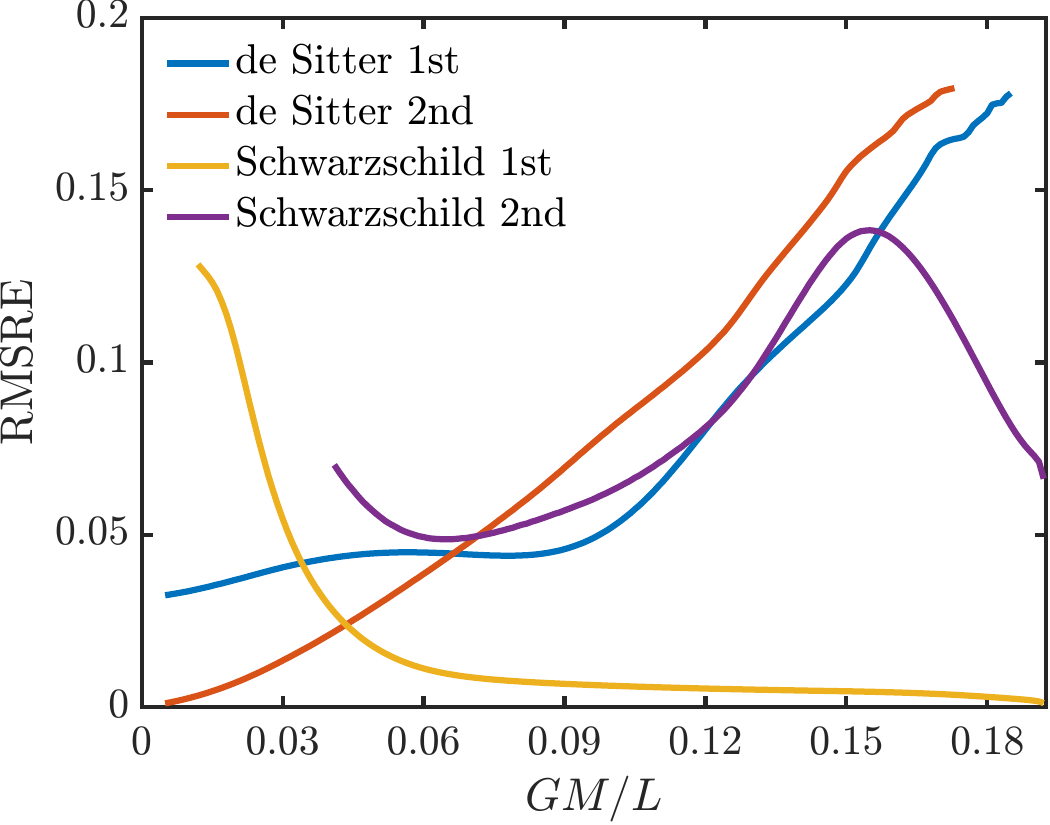}
\caption{Plot of root mean squared relative error for fitting a single QNM to ultra slow-roll data for various $GM/L$, for the first two modes from each branch. For $GM/L \lesssim 0.043$, the 2nd de Sitter mode gives the best single mode fit, and then for $GM/L \gtrsim 0.043$, the 1st Schwarzschild mode is preferred.}
\label{SingleRMSREPlot}
\end{figure}

The results are shown in Figure \ref{SingleRMSREPlot}, where we plot the RMSRE of the single mode fits, for a range of $GM/L$, to get a measure of which single mode gives the best fit. To collect data for these fits, we include a sample of 50 evenly spaced radii over the grid, and $3 \times 10^3$ timesteps, starting for each radius when the curve reaches 50\% of its eventual maximum height, to allow the non-linearities of the transition to dissipate. Nevertheless, we observe in Figure \ref{USRfits} that the fit tends to also match the data outside of the range of timesteps fitted. We use a range of $GM/L$ from $0.005$ to $0.192$, in steps of $0.001$. Each branch only appears in this plot for some subset of the range of $GM/L$ considered, since near the edges of $GM/L$ some modes are challenging to numerically resolve by QNMSpectral (as seen from the rapidly decreasing imaginary parts of some branches at the edges in Figure \ref{ImagFreqPlot}).
We see that for $GM/L \lesssim 0.043$, the 2nd de Sitter mode is the best single mode fit, and for $GM/L \gtrsim 0.043$, the 1st Schwarzschild mode is the best single mode fit. If we consider only the single mode fit with the lowest RMSRE for each $GM/L$, then the errors of the resulting fits appear to be lower in the well-separated horizons and near-Nariai limits, and higher in the intermediate range of $GM/L$. The maximum error of these single mode fits is \mbox{$\approx$ 2.4\%}.

Since the mode with the lowest single mode fit RMSRE appears to change as we vary $GM/L$, we can also consider a two mode fit to model this progression. For some intermediate range of $GM/L$, we include both the 2nd de Sitter mode (preferred near pure de Sitter), and the 1st Schwarzschild mode (preferred near Nariai). Since both of these modes cannot be easily resolved by the QNMSpectral results near the edges of $GM/L$, we will necessarily be restricted to a one mode fit there. Further, due to the overfitting cautions of \cite{Giesler:2019uxc, Zhu:2023mzv, Baibhav:2023clw}, even if we could use a two mode fit over the whole range of $GM/L$, we would likely find that the additional unnecessary freedom in the fits would overfit the noise.
As such, in the main plots for the combined mode fits in Figures \ref{CombinedNormAmpsPlot} and \ref{CombinedRMSREPlot}, we consider a one mode fit for $GM/L < 0.012$ (2nd de Sitter) and for $GM/L > 0.083$ (1st Schwarzschild), and a two mode fit for $0.012 \leq GM/L \leq 0.083 $. These cutoff values are chosen to be the first time that the negligible mode has an amplitude with $|A_{\textrm{Sch/dS}}|^2 < 10^{-7}$, as the edges of $GM/L$ are approached. If we pushed the cutoffs much closer to the edges of $GM/L$, we would risk the insignificant mode overfitting numerical noise. Note that since both the amplitudes in Figure \ref{CombinedNormAmpsPlot} and the RMSRE in Figure \ref{CombinedRMSREPlot} appear continuous at the cutoff between single and combined mode fits (up to numerical precision), this does suggest that those results are consistent, despite fitting with a different number of modes on each side of the cutoffs.
This concludes the formative rationale leading to the combined mode fit in Section \ref{FittingSec}.

\section{Useful coordinate transformations} \label{Penroseapp}

In this appendix, we give details of various coordinate transformations that we discuss throughout the work.

In cosmological spacetimes, we often work in the FRW coordinates of \eqref{FRWMetric}. However, in black hole spacetimes, we instead often work in \textit{static} coordinates, where the metric components do not explicitly depend on a time coordinate. 
For pure de Sitter, we can change coordinates to a static frame $(t, r)$ through the following transformation,
\begin{equation} \label{dScoordstrafo}
    T_c = t + \frac{1}{2H} \log(1 - H^2 r^2), \quad R_c = r e^{-HT_c} = \frac{r}{\sqrt{1-H^2 r^2}} e^{-Ht},
\end{equation}
giving the de Sitter metric in \eqref{BackgroundSdSMetric}, via the special case $M=0$.

The static time $t$ of the SdS metric in \eqref{BackgroundSdSMetric} is singular at the horizons, however we can consider Eddington-Finkelstein (EF) coordinates to cure this issue.
Recall the SdS metric given in \eqref{BackgroundSdSMetric} in terms of $f(r)$ in \eqref{Backgroundfr}. The horizons are determined through $f(r) = 0$, and provided $M, \Lambda > 0$, the resulting effective cubic equation can be factorised as,
\begin{equation} \label{frcubic}
    -\frac{H^2}{r} (r - r_b)(r - r_c)(r - r_n) = 0.
\end{equation}
For $GM \sqrt{\Lambda} < 1/3$, this equation has three real solutions, with $r_b < r_c$, corresponding to the black hole horizon and cosmological horizon respectively, and $r_n = -(r_b + r_c)$ (the unphysical negative root).
We can introduce the tortoise function $r^*$,
\begin{equation} \label{Tortoisedef}
    \frac{dr^{*}}{dr} = \frac{1}{f(r)}.
\end{equation}
Then in terms of the tortoise $r^*$, the EF coordinates are,
\begin{equation} \label{EFdef}
    v = t + r^*, \quad u = t - r^*,
\end{equation}
which give notions of time that are regular at the black hole and cosmological horizons, respectively.
The tortoise coordinate $r^*$ can be obtained by integrating the differential equation $\eqref{Tortoisedef}$, and using the factorisation of $f(r)$ given in \eqref{frcubic}. As such, there is an ambiguity of the constant of integration in the definition of $r^*$. This ambiguity feeds in to the definition of the EF coordinates, and the resulting Kruskal coordinates (but leads only to a rescaling). 
In this work, we choose the constant of integration as,
\begin{equation} \label{TortoiseFullExpr}
    r^*(r) = \int \frac{dr}{f(r)} = \frac{1}{2\kappa_c} \log{\left|r_c - r\right|} + \frac{1}{2\kappa_b} \log{\left|r - r_b\right|} + \frac{1}{2\kappa_n} \log{\left|r - r_n\right|},
\end{equation}
where $\kappa_i$ are the surface gravities (which can be related to temperatures in thermodynamic horizon first laws \cite{Kubiznak:2016qmn}), defined by $2 \kappa_i = f'(r_i)$.
Since we set $L = 1$ in much of our calculations, this makes lengths dimensionless, so that the logarithms above are well-defined\footnote{In general, this choice amounts to instead using $|r-r_h|/L$ in each logarithm.}.

To construct Penrose diagrams, we can identify a choice of compactification ``time'' and ``radial'' coordinate,
\begin{equation} \label{TimeSpacePenrose}
\begin{aligned}
    \hat{T} & = \frac{1}{2} \left[\tanh\left({\frac{|\kappa_c| v}{2}}\right) + \tanh\left({\frac{|\kappa_c| u}{2}}\right) \right] \\ \hat{R} & = \frac{1}{2} \left[\tanh\left({\frac{|\kappa_c| v}{2}}\right) - \tanh\left({\frac{|\kappa_c| u}{2}}\right) \right],
\end{aligned}
\end{equation}
which we use for the Penrose diagrams in Figures \ref{fig:grid_penrose}, \ref{QNMcompareplot} and \ref{QNMNHplot}.

For each horizon, one can generate Kruskal coordinates that can be used to extend over the horizon \cite{Poisson:2009pwt}. We can consider the following cosmological Kruskals\footnote{Technically these are only correct definitions of $U$ and $V$ in the static patch, $V$ has to be analytically extended to change sign over the horizon, see for example \cite{Poisson:2009pwt}.},
\begin{equation} \label{KruskalCosmoDef}
    U \equiv \frac{1}{|\kappa_c|} e^{|\kappa_c|u}, \quad V \equiv - \frac{1}{|\kappa_c|} e^{-|\kappa_c|v}.
\end{equation}

For the near cosmological horizon contour plot in Figure \ref{QNMNHplot}, we extend the static patch near horizon result \eqref{StaticPatchQNMNH} using the Kruskals. Multiplying $U$ and $V$ gives a purely radial function in the static patch,
\begin{equation} \label{UVrmap}
    - |\kappa_c|^2 UV = (r_c - r)(r - r_b)^{-\frac{|\kappa_c|}{\kappa_b}}(r - r_n)^{-\frac{|\kappa_c|}{\kappa_n}}.
\end{equation}
Therefore taking $U, V \in (-\infty, \infty)$ provides an analytic continuation of $r$ from $(r_b, r_c)$ to $(r_b, \infty)$, which we can numerically invert for the near cosmological horizon expansion of \eqref{StaticPatchQNMNH}.

\bibliography{bibli.bib}
\bibliographystyle{unsrt}

\end{document}